\documentclass[aps, prd, twocolumn, superscriptaddress, nofootinbib, showpacs,preprintnumbers]{revtex4}
\usepackage[dvipdfmx]{graphicx}
\usepackage{amsmath,amssymb}
\usepackage{mathrsfs}
\usepackage{bm}
\usepackage{booktabs}
\usepackage{comment}

\def\fsl#1{\setbox0=\hbox{$#1$}                 % set a box for #1 
   \dimen0=\wd0                                 % and get its size
   \setbox1=\hbox{/} \dimen1=\wd1               % get size of /
   \ifdim\dimen0>\dimen1                        % #1 is bigger
      \rlap{\hbox to \dimen0{\hfil/\hfil}}      % so center / in box
      #1                                        % and print #1
   \else                                        % / is bigger
      \rlap{\hbox to \dimen1{\hfil$#1$\hfil}}   % so center #1
      /                                         % and print /
   \fi}                                         %
\newcommand{\tr}{\mbox{tr}}

\newcommand{\VEV}[1]{\langle #1 \rangle}
\makeatletter
\@addtoreset{equation}{section}
\makeatother

\begin{document}
\preprint{MISC-2016-08}
\title{
Relic Abundance in a Secluded Dark Matter Scenario with a Massive Mediator
}  
\date{\today}

\author{Shohei Okawa}
\email[E-mail: ]{okawa@eken.phys.nagoya-u.ac.jp}
\affiliation{
  Department of Physics, Nagoya University, Nagoya 464-8602, Japan
}
\author{Masaharu Tanabashi}
\email[E-mail: ]{tanabash@eken.phys.nagoya-u.ac.jp}
\affiliation{
  Department of Physics, Nagoya University, Nagoya 464-8602, Japan
}
\affiliation{
  Kobayashi-Maskawa Institute for the Origin of Particles and the Universe, 
  Nagoya University, Nagoya 464-8602, Japan
}
\author{Masato Yamanaka}
\email[E-mail: ]{masato.yamanaka@cc.kyoto-su.ac.jp}
\affiliation{
  Maskawa Institute, Kyoto Sangyo University, Kyoto 603-8555, Japan
}

\begin{abstract}
The relic abundance of the dark matter (DM) particle $d$
is studied in a secluded DM scenario, in which the $d$ number
decreasing process dominantly occurs not through the pair annihilation 
of $d$ into the standard model particles, but via the $dd \to mm$ 
scattering process with a subsequently decaying mediator particle $m$.
It is pointed out that the cosmologically observed relic abundance of DM
can be accomplished even with a massive mediator having a mass $m_m$ 
non-negligibly heavy compared with the DM particle mass $m_d$.
In the degenerated $d$-$m$ case 
($m_d=m_m$), 
the DM relic abundance is realized by adjusting 
the $dd \to mm$ scattering amplitude large enough and by choosing
an appropriate mediator particle life-time.
The DM evolution in the early universe exhibits characteristic
``terrace'' behavior, or two-step number density decreasing behavior,
having a ``fake'' freeze-out at the first step.
Based on these observations, a novel possibility of the DM model 
buildings is introduced
in which the mediator particle $m$ is unified with the DM particle 
$d$ in an approximate dark symmetry multiplet.
A pionic DM model is proposed to illustrate this idea 
in a renormalizable field theory framework.

\end{abstract}

\maketitle

\section{Introduction}
\label{sec:introduction}
More than 80\% of the matter is made up 
of Dark Matter (DM) in the universe\cite{Agashe:2014kda,Hinshaw:2012aka,Ade:2015xua}.
Very little of the DM nature is known, however, besides
its cosmological abundance $\Omega_{\rm dm} h^2=0.1188 \pm 0.0010$~\cite{Ade:2015xua}.

Theories beyond the standard model (BSM) in particle physics 
often predict the existence of new massive particles.
The lightest neutral stable new particle, if it exists in these BSM 
scenarios, provides an excellent DM candidate, 
longevity of which is guaranteed by a new symmetry existing
in the BSM scenario.
For reviews, see, e.g., Refs.\cite{Bertone:2004pz,Feng:2010gw}.

A promising hypothesis we are able to make in these DM models 
is that the DM particles were produced thermally 
in the early universe (thermal relic hypothesis)\cite{Lee:1977ua}
and the cosmological DM abundance can be
computed through the Boltzmann equation, once the mass
and the couplings of the DM particle are fixed.
It has been widely assumed that the DM number density 
decreasing process was dominated by the pair annihilation 
of the DM particles into the standard model (SM) particles.
If this is true, the DM particle mass and its couplings with the
SM particles can be related with each other through
the observed value of the cosmological DM abundance.
Assuming further the DM pair annihilation cross section is determined
by the couplings of the order of the electroweak gauge coupling 
strengths, the thermal relic hypothesis predicts the DM particle
mass of the weak scale.
This striking coincidence is often called the ``WIMP Miracle''.
The heavier DM particle mass,
the stronger DM couplings with the SM sector we need to assume in 
this widely adopted thermal relic scenarios:
we therefore cannot seclude the DM particles from the SM sector.
This tendency has encouraged the DM particle searches in the direct 
detection experiments~\cite{Undagoitia:2015gya}
and in the collider experiments~\cite{Abercrombie:2015wmb}.
It is unfortunate, however, up to the present time,
we have no fully confidently
positive results in these DM particle search 
experiments~\cite{Undagoitia:2015gya,Tan:2016zwf,Manalaysay2016,Akerib:2016vxi,Agnes:2015ftt,Agnese:2015ywx}. 

Recently, new varieties of thermal relic DM scenarios have been proposed.
In these scenarios, interactions between the DM particles 
and the SM particles are weak enough to make the scenarios 
consistent with the present and near future constraints from the 
direct DM detections and the collider experiments, 
still keeping the observed value of the cosmological
DM abundance,
by introducing novel mechanisms to decrease the 
DM particle number density in the thermal history of the early
universe.

The authors of Refs.\cite{Hochberg:2014dra,Hochberg:2014kqa} 
consider a scenario in which the relic density
is controlled by the $3\to 2$ scattering process among the DM particles,
instead of the traditional $2\to 2$ process of the DM pair annihilation 
into the SM particles. 
The cross section for the $3\to 2$ scattering enough to explain 
the observed DM abundance may be achieved in the DM model 
associated with dark strong dynamics, assuming the dark pions to be 
the DM particle.  
The anomaly induced Wess-Zumino-Witten (WZW) 
term\cite{Wess:1971yu,Witten:1983tw}
in the dark chiral Lagrangian
naturally explains the required $3\to 2$ scattering in the dark sector.
Although the dark sector in this scenario is chemically secluded from the
visible sector, it is assumed to interact with the SM sector very weakly,
keeping the dark sector temperature equal to the visible sector
in the epoch of the DM particle number decreasing processes.
It has been pointed out, however, the WZW induced 
$3\to 2$ scattering annihilation process may not be enough to reduce the 
DM number density if the higher order effects are incorporated consistently
in the chiral perturbation framework.

Refs.\cite{Pospelov:2007mp,ArkaniHamed:2008qn} 
introduce a phenomenological scenario in which 
the DM number density decreasing mechanism is implemented by 
the DM pair annihilation into additional non-SM particles (mediator
particles) through the $2\to 2$ process (the secluded DM scenario).
The mediator is assumed to decay into the SM particles later and
is sufficiently lighter than the DM particle.
Note here the mass hierarchy between the two separate mass scales
(the DM mass $m_d$ and the mediator mass $m_m$), 
$m_d \gg m_m$, is introduced on an ad hoc basis
in this scenario.  
With a large mass gap $m_d \gg m_m$,
the DM relic abundance is insensitive to the mediator particle 
$m$ life-time.
It is assumed that the dark sector
decouples chemically from the visible sector in the early epoch
of its thermal history, 
but it follows the visible sector temperature,
keeping the kinematical equilibrium with the SM particles.

The authors of Ref.\cite{Pappadopulo:2016pkp} propose a DM sector
almost completely decoupled from the SM sector both 
kinematically and chemically (cannibal DM).
Novel mechanisms similar to SIMP and the secluded DM 
are responsible for the DM particle number
decreasing process in the dark sector.
Since the dark sector is decoupled from the SM sector almost completely, 
the DM particle
temperature differs from the SM sector in the thermal history
of the universe.
Ref.\cite{Berlin:2016vnh} considers a similar scenario having the
dark sector kinematically decoupled from the visible sector.  The late
time decay of new particles into the SM sector, which were produced 
copiously in the dark sector thermal history, reheats the visible 
sector and dilutes the DM density.

In this paper,
we solve the Boltzmann equation numerically in a 
toy secluded DM model having a heavy mediator $m_d \sim m_m$. 
The secluded DM having a large mass gap 
$m_d \gg m_m$,
as well as the familiar mechanism of the DM pair annihilation into the
SM particles,
can also be analyzed in this toy model.
We point out the seclusion mechanism works well even 
with $m_d \sim m_m$ if the mediator life-time is short enough,
in contrast to the original secluded DM proposal having
a large mass gap $m_d \gg m_m$ and a longer life-time mediator.
The hierarchy assumption $m_d \gg m_m$ 
made ad hoc in Refs.\cite{Pospelov:2007mp,ArkaniHamed:2008qn} 
is, therefore, not necessary.
We notice that the departure of the mediator number density 
from its thermoequilibrium value causes non-negligible 
effects in the evaluation of the DM relic density.

Especially, in the $m_d=m_m$ case with sufficiently strong
$d$-$m$ interaction, the evolutions of the DM density exhibits 
interesting behavior, i.e., two step DM density decreasing. 
At the first step, the dark sector chemically decouples
from the SM particles, and the DM density is temporarily 
frozen to a value much higher than usual thermal relic DM 
scenario (``fake'' freeze-out).
This is ``fake'' because the DM particle still interacts
with the mediator particle strongly, and it keeps the 
chemical equilibrium with the mediator.
The next step DM density decreasing starts when the mediator
decay becomes active.
The relic abundance of the DM density is therefore controlled by 
the mediator life-time.
The true freeze-out takes place only after the DM particle 
decouples from the mediator.
The evolution of the DM density thus exhibits characteristic
``terrace'' structure in this setup as shown in Fig.~\ref{Fig:evolution_1} 
later.

The realization of the secluded mechanism with $m_d \sim m_m$ 
opens a new fascinating window on the DM model buildings,
allowing unified descriptions for the mediator and the DM in the 
secluded scenarios.
We give a concrete example of unified description of mediator
and DM, in which
both mediator and DM particles are realized as the pseudo Nambu-Goldstone
bosons, in a manner similar to the
SIMP scenario.
We emphasize that the key process in the secluded DM scenario, i.e., 
the $2\to 2$ process 
of the DM pair annihilation into mediator particles, is provided by the 
low energy theorem amplitude and is well under theoretical control
in our concrete model.

This paper is organized as follows:
In Sec.~\ref{sec:toy-model-secluded}, we give a toy model
of the secluded DM model.  
A brief review on the Boltzmann equation we use in our
numerical analysis is also described there.
The results on the numerical computations on the evolutions
of the DM and mediator number densities are presented
in Sec.~\ref{sec:numerical-results}.
We then propose a pionic DM scenario based on our numerical
computation in Sec.~\ref{sec:real-pion-dark}.
Sec.~\ref{sec:summary} is devoted for summary and outlook.
We ensure comprehension of the numerical evolutions 
by analytically describing the relevant quantities in
Appendix.~\ref{Sec:ana_est}.

\section{
Boltzmann equation in A Toy Model for Secluded Dark Matter}
\label{sec:toy-model-secluded}

It is possible to write down a toy model in which 
several known DM density decreasing mechanisms, such as 
the familiar thermal relic DM, the cannibal DM
with a decaying mediator, and the secluded DM with a mass gap 
$m_d \gg m_m$, are described in a unified manner.
The secluded DM scenario with $m_d \sim m_m$ we consider in this paper
can also be 
implemented in this toy model as well.
We here introduce such a toy model and
write down the coupled Boltzmann equations 
governing the evolutions of the DM and the mediator number densities. 

\subsection{A toy model} 
\label{Sec:toy}

We introduce two real scalar fields $\phi_d$ and $\phi_m$, which correspond
to the DM and mediator particles $d$ and $m$, respectively.
They interact with each other and also with the SM weak doublet Higgs 
field $\phi$.
The model is described by the Lagrangian, 
\begin{equation}
\begin{split}
	\mathcal{L}_{toy} 
	=& 
        \dfrac{1}{2} (\partial_\mu \phi_d)^2 - \dfrac{1}{2} m_d^2 \phi_d^2
        \\&+
        \dfrac{1}{2} (\partial_\mu \phi_m)^2 - \dfrac{1}{2} m_m^2 \phi_m^2
        \\&+
	g_{ddmm} 
	\bigl( \phi_{d} \phi_{d} \bigr) 
	\bigl( \phi_{m} \phi_{m} \bigr)
	+
	g_{m\phi^\dagger\phi} m_{m} \phi_{m} \phi^{\dagger} \phi
	\\&+ 
	g_{dd\phi^\dagger\phi} \phi_{d} \phi_{d} \phi^{\dagger} \phi 
	\\&+ 
	g_{mm\phi^\dagger\phi} \phi_{m} \phi_{m} \phi^{\dagger} \phi, 
	\label{Eq:Lag_sce1}
\end{split}
\end{equation}
with $m_{d}$, $m_{m}$  being masses of the DM particle $d$
and the mediator particle $m$, respectively.
The longevity of the DM particle $d$ is protected by the $Z_2$ symmetry 
$\phi_d \leftrightarrow -\phi_d$.

We know several parameter regions 
explain the
cosmologically observed DM relic abundance in the present toy model.
\begin{itemize}
\item 
  If we take the $m_m$ much heavier than the DM mass,
  \begin{displaymath}
    m_m \gg m_d, 
  \end{displaymath}
  we can integrate out the $m$ particle in the Lagrangian and
  we only have two parameters $m_d$ and $g_{dd\phi^\dagger\phi}$,
  which can be chosen to obtain the relic abundance.
  This model is nothing but the conventional Higgs portal DM~\cite{Silveira:1985rk,McDonald:1993ex,Burgess:2000yq},
  a typical scenario in the familiar thermal relic DM.
\item If we take 
  \begin{displaymath}
    g_{dd\phi^\dagger\phi}=g_{mm\phi^\dagger\phi}=0,
  \end{displaymath}
  and very small $g_{m\phi^\dagger\phi}$, 
  the dark sector particles $d$ and $m$ decouple from the visible sector
almost completely and the thermal equilibrium with the visible sector 
is lost.
For 
\begin{displaymath}
   m_d \gg m_m,
\end{displaymath}
the mediator $m$ decays into the visible particles 
after the decoupling between $m$ and $d$,
due to the long life-time of $m$,
$\propto 1/(m_m g_{m\phi^\dagger\phi}^2)$. 
In this case,  $g_{ddmm}$ and $m_d$ can be chosen to obtain the observed
relic density.
This scenario can be considered as a Higgs portal realization of the
cannibal DM model~\cite{Pappadopulo:2016pkp}.
\item 
It is also possible to consider a scenario in which both 
$g_{dd\phi^\dagger\phi}$ and $g_{mm\phi^\dagger\phi}$ are non-zero but satisfy
  \begin{displaymath}
    g_{dd\phi^\dagger\phi} \ll 1, \qquad
    g_{mm\phi^\dagger\phi} \ll 1 . 
  \end{displaymath}
The DM coupling strengths are arranged so as the dark sector to keep
its thermal contact with the visible sector even after its chemical decoupling.
Similarly to the cannibal DM case, if we take  
\begin{displaymath}
 m_d \gg m_m 
\end{displaymath}
and non-vanishing $g_{m\phi^\dagger\phi}$,
the DM density decreasing process takes place in the $dd \to mm$ scattering
process. 
The coupling $g_{ddmm}$ 
and the DM mass $m_d$ control the relic abundance.
This possibility (the secluded DM with a large mass gap) 
has been known since Refs.~\cite{Pospelov:2007mp,ArkaniHamed:2008qn}.
\end{itemize}

Note that the $g_{dd\phi^\dagger\phi} \ll g_{mm\phi^\dagger\phi}$ model 
with $g_{mm\phi^\dagger\phi} \sim 0.1$ and $g_{ddmm} \sim 0.1$ can also
accommodate the appropriate DM relic density.
See Ref.\cite{Belanger:2011ww} for a study of this possibility in the 
stable mediator limit.
It will be dealt with further in a separate publication.

In the following sections, we consider yet another realization 
to obtain the appropriate relic abundance by choosing
the life-time of the mediator particle $m$ in
a novel parameter region $m_m \sim m_d$ and $g_{dd\phi^\dagger\phi}\ll 1$,
$g_{mm\phi^\dagger\phi}\ll 1$,
which guarantee the thermal equilibrium between the dark sector and SM fields
in the epoch of its chemical decoupling from the SM 
particles.

\subsection{Boltzmann equation with a species going out of equilibrium} 
\label{Sec:Boltz}

Here we describe a procedure to obtain 
the Boltzmann equation for a particle $i$ valid even when
out-of-chemical-equilibrium particles $j$, $X$ and $Y$ 
are interacting with the particle $i$.
We restrict ourselves to the case in which all of these
particles keep kinematical equilibriums with the thermal bath and
feel the same temperature $T$.
The validity of this assumption in our DM relic density analysis 
will be discussed later in Sec.~\ref{Sec:T}.

We illustrate the procedure in a simple setup in which 
only two processes, 
(i) decay and inverse decay $i \leftrightarrow XY$, 
(ii) scattering process $ij \leftrightarrow XY$,
are responsible.
Using Eq.~(5.11) in Ref.~\cite{Kolb:1990vq},
the Boltzmann equation for $i$ is given by
\begin{equation}
\begin{split}
	\frac{dn_{i}}{dt} + 3H n_{i} 
	=& 
	- \int \hspace{-1pt} d\Pi_{i} d\Pi_{X} d\Pi_{Y} 
	\\& \hspace{3mm} \times 
	(2\pi)^{4} \delta^{(4)} (p_{i} - p_{X} - p_{Y})
	\\& \hspace{3mm} \times 
	|\mathcal{M}|_{i \leftrightarrow  XY}^{2} 
	\bigl( f_{i} - f_{X} f_{Y} \bigr)
	\\&
	- \int \hspace{-1pt} d\Pi_{i} d\Pi_{j} d\Pi_{X} d\Pi_{Y} 
	\\& \hspace{3mm} \times 
	(2\pi)^{4} \delta^{(4)} (p_{i} + p_{j} - p_{X} - p_{Y})
	\\& \hspace{3mm} \times 
	|\mathcal{M}|_{ij \leftrightarrow  XY}^{2} 
	\bigl( f_{i} f_{j} - f_{X} f_{Y} \bigr), 
\label{eq:boltzmann1}
\end{split}     
\end{equation}
with  
\begin{equation}
d\Pi_{a} = \dfrac{g_{a}}{(2\pi)^{3}} \dfrac{d^{3} 
\boldsymbol{p_a}}{2E_{a}}
\end{equation}
denoting a Lorentz invariant phase space for $a=i,j,X,Y$. 
Here $g_a$ stands for the internal degree of freedom for 
particle $a$.
The Hubble rate $H$ is given by
\begin{equation}
  H = 1.66 g_*^{1/2} \dfrac{T^2}{M_{\rm pl}},
\end{equation}
where $g_*$ represents the total number of relativistic degrees
of freedom for particles.
We use $g_*=106.75$ to simplify the numerical analysis 
throughout the present paper.  
The Planck mass is denoted by $M_{\rm pl}$.

In order to perform the phase space integrals in Eq.(\ref{eq:boltzmann1}),
we assume the distribution functions $f_a$ are approximately given 
by the Maxwell-Boltzmann distribution form, $f_{a} = 
e^{-(E_{a} - \mu_{a})/T}$.
Here $\mu_a$ represents the value of the chemical potential for $a$.
The $\delta$-functions enforce 
$E_{i} = E_{X} + E_{Y}$ and $E_{i} + E_{j} = E_{X} + E_{Y}$. 
Then the distribution functions are rewritten as 
$f_{i} - f_{X} f_{Y} = e^{-E_{i}/T} (e^{\mu_{i}/T} - 
e^{(\mu_{X} + \mu_{Y})/T})$, and $f_{i} f_{j} - f_{X} 
f_{Y} = e^{-(E_{i} + E_{j})/T} (e^{(\mu_{i} + \mu_{j})/T} 
- e^{(\mu_{X} + \mu_{Y})/T})$. 
Note that the number density $n_{a}^{eq}$ for a species $a$
in chemical equilibrium with the thermal bath is given by
\begin{displaymath}
  n_{a}^{eq} = \dfrac{g_{a}}{(2 \pi)^{3}} \int \hspace{-1pt} d^{3} 
  \boldsymbol{p}_{a} e^{-E_{a}/T} .
\end{displaymath}
The actual number density $n_{a}$ for a species $a$ out of
chemical equilibrium 
is related with the chemical potential $\mu_a$ as
$n_{a} = e^{\mu_{a}/T} n_{a}^{eq}$.

We introduce thermally averaged decay rates and thermally averaged 
cross sections as follows, 
\begin{equation}
\begin{split}
	\langle \Gamma \rangle_{i \leftrightarrow XY} 
	&= 
	\frac{1}{n_{i}^{eq}}
	\int \hspace{-1pt} d\Pi_{i} d\Pi_{X} d\Pi_{Y} 
	\\& \hspace{3mm} \times
	(2\pi)^{4} \delta^{(4)} (p_{i} - p_{X} - p_{Y})
	\\& \hspace{3mm} \times 
	|\mathcal{M}|_{i \leftrightarrow XY}^{2} 
	e^{-E_{i}/T}, 
	\label{Eq:thermalGamma_1}
\end{split}     
\end{equation}
%%%
%%%
\begin{equation}
\begin{split}
	\langle \sigma v \rangle_{ij \leftrightarrow XY} 
	&= 
	\frac{1}{n_{i}^{eq} n_{j}^{eq}}
	\int \hspace{-1pt} d\Pi_{i} d\Pi_{j} d\Pi_{X} d\Pi_{Y} 
	\\& \hspace{3mm} \times
	(2\pi)^{4} \delta^{(4)} (p_{i} + p_{j} - p_{X} - p_{Y})
	\\& \hspace{3mm} \times 
	|\mathcal{M}|_{ij \leftrightarrow XY}^{2} 
	e^{-(E_{i} + E_{j})/T},
	\label{Eq:22sigmav_1}
\end{split}     
\end{equation}
where $v$ is relative velocity of initial particles. 
The phase space integrals in the Boltzmann equation Eq.(\ref{eq:boltzmann1})
can now be performed.  We obtain
\begin{equation}
\begin{split}
	\frac{dn_{i}}{dt} + 3H n_{i} 
	=& 
	- \left\{ 
	e^{\mu_{i}/T} - e^{(\mu_{X} + \mu_{Y})/T} 
	\right\} 
	n_{i}^{eq} 
	\langle \Gamma \rangle_{i \leftrightarrow XY}
	\\&
	- \left\{ 
	e^{(\mu_{i} + \mu_{j})/T} - e^{(\mu_{X} + \mu_{Y})/T} 
	\right\} 
	n_{i}^{eq} n_{j}^{eq} 
	\\& \hspace{3mm} \times 
	\langle \sigma v \rangle_{ij \leftrightarrow XY}
	%%%
	%%%
	\\=& 
	- \left\{ 
	n_{i} - n_{i}^{eq}  
	\frac{n_{X}}{n_{X}^{eq}} 
	\frac{n_{Y}}{n_{Y}^{eq}}
	\right\} 
	\langle \Gamma \rangle_{i \leftrightarrow XY}
	\\&
	- \left\{ 
	n_{i} n_{j} - n_{i}^{eq} n_{j}^{eq} 
	\frac{n_{X}}{n_{X}^{eq}} 
	\frac{n_{Y}}{n_{Y}^{eq}}
	\right\} 
	\langle \sigma v \rangle_{ij \leftrightarrow XY}, 
	\label{Eq:Boltz_2}
\end{split}     
\end{equation}
which is expressed in terms of the actual number density $n_a$ and 
the number density in chemical equilibrium $n_a^{eq}$.
In the case where species $X$ and $Y$ are in chemical 
equilibrium with the thermal bath, $n_{X(Y)}/n_{X(Y)}^{eq} = 1$, 
the Boltzmann equation Eq.(\ref{Eq:Boltz_2}) reduces to a 
``familiar'' form.
Once species $X$ and/or $Y$ go out of equilibrium, their 
number densities deviate from equilibrium values, i.e., 
$n_{X(Y)}/n_{X(Y)}^{eq} \neq 1$, which makes an important
difference from the analysis based on the familiar Boltzmann equation.
The non-unity ratio
$n_{X(Y)}/n_{X(Y)}^{eq} \neq 1$ can trigger non-equilibration 
of species $i$ even if the rates of the 
processes $i\leftrightarrow XY$ and $ij \leftrightarrow XY$ are
faster than the Hubble rate $H$.

The assumption we made on the distribution functions $f_a$ cannot be
justified if the departures from their chemical and thermal equilibriums are 
large.
It should be noted, however, the validity of this approximation can be 
guaranteed in the situation in which all of particles $i$, $j$, $X$ and 
$Y$ start to deviate from their chemical equilibriums almost simultaneously.

\subsection{Boltzmann equation for the DM-mediator system}
\label{Sec:Bol_sys}  

We are now ready to derive the Boltzmann equations which determine
the evolutions of the number densities of the DM particle $d$ and 
the mediator particle $m$ in our toy model Eq.(\ref{Eq:Lag_sce1}).
The number changing processes of the DM and the mediator 
are
\begin{equation}
\begin{split}
	m &\leftrightarrow \phi^{\dagger} \phi, 
	\\
	dd (mm) &\leftrightarrow \phi^{\dagger} \phi, 
	\\
	dd &\leftrightarrow mm. 
	\label{Eq:DM_Med_process_toy}
\end{split}
\end{equation}
The interplay between these processes determines the DM relic density. 

The Boltzmann equations are derived by implementing the
processes~\eqref{Eq:DM_Med_process_toy} 
in \eqref{Eq:Boltz_2},
\begin{equation}
\begin{split}
	\frac{dn_{d}}{dt} + 3Hn_{d}
	=&- 
	\langle \sigma v \rangle_{dd 
	\leftrightarrow \phi^{\dagger} \phi}
	\Bigl[ 
	n_{d}^{2} - \bigl( n_{d}^{eq} \bigr)^{2}
	\Bigr]
	\\& \hspace{-13mm} - 
	\langle \sigma v \rangle_{dd 
	\leftrightarrow mm} 
	\biggl[ 
	n_{d}^{2} - \bigl( n_{d}^{eq} \bigr)^{2}
	\frac{n_{m}^{2}}{\bigl( n_{m}^{eq} \bigr)^{2}}
	\biggr], 
	\label{Eq:Boltz_1_1}
\end{split}
\end{equation}
%%%
%%%
\begin{equation}
\begin{split}
	\frac{dn_{m}}{dt} + 3Hn_{m}
	=& 
	- \langle \Gamma \rangle_{m \leftrightarrow \phi^{\dagger} \phi}
	\Bigl[ 
	n_{m} - n_{m}^{eq}
	\Bigr]
	\\& \hspace{-13mm} - 
	\langle \sigma v \rangle_{mm 
	\leftrightarrow \phi^{\dagger} \phi}
	\Bigl[ 
	n_{m}^{2} - \bigl( n_{m}^{eq} \bigr)^{2}
	\Bigr]
	\\& \hspace{-13mm} -  
	\langle \sigma v \rangle_{mm \leftrightarrow dd} 
	\biggl[ 
	n_{m}^{2} - \bigl( n_{m}^{eq} \bigr)^{2}
	\frac{n_{d}^{2}}{\bigl( n_{d}^{eq} \bigr)^{2}}
	\biggr] . 
	\label{Eq:Boltz_1_2}
\end{split}
\end{equation}
We assumed that the SM Higgs is in chemical equilibrium, i.e, 
$n_{\phi} = n_{\phi}^{eq}$. 

Note that, if the mediator couples with the SM particles
sizably via $g_{m\phi^\dagger\phi}$ or $g_{mm\phi^\dagger\phi}$,
the mediator keeps its chemical equilibrium with the SM Higgs 
through the decay and inverse decay process $m \leftrightarrow \phi \phi^\dagger$, 
or through the $mm \leftrightarrow \phi \phi^\dagger$ process.
If the chemical equilibrium of the mediator particle lasts 
until the final DM freeze-out epoch, 
the mediator can be regarded as a part of background thermal plasma
in the $dd\leftrightarrow mm$ process.
In this case, the DM relic density is determined almost solely by $g_{ddmm}$
and becomes insensitive to the values of $g_{m\phi^\dagger\phi}$ and 
$g_{mm\phi^\dagger\phi}$.

The situation becomes a bit elaborate, 
if $m$ goes out of chemical equilibrium before 
$d$ decouples from the 
mediator $m$.
All of couplings $g_{ddmm}$, $g_{m\phi^\dagger\phi}$ and
$g_{mm\phi^\dagger\phi}$ are equally important in the determination
of the DM relic abundance in this case.

In the remaining of this section, we analytically derive the critical 
value of the $g_{m\phi^\dagger\phi}$ coupling, which separates these two phases.

We include only the $m$ (inverse) decay process in the Boltzmann equation
(\ref{Eq:Boltz_1_2}) as the 
reaction between the mediator and the SM fields.
This simplification is reasonable, because the decay 
dominates over the scattering with the SM Higgs when the 
temperature drops below $m$ mass. 
We introduce a variable $X_{m} = n_{m} R^{3}$, where $R$ 
denotes the scale factor of the universe. The Boltzmann 
equation of $m$ can be rewritten in terms of $X_{m}$, 
\begin{equation}
\begin{split}
	\frac{dX_{m}}{dt} 
	= 
	- \frac{1}{2}\langle \Gamma \rangle_{m 
	\leftrightarrow \phi^{\dagger} \phi} 
	\left( X_{m} - X_{m}^{eq} \right), 
	\label{Eq:Boltz_Xm}
\end{split}     
\end{equation}
where we assumed $m$ and $d$ keeps their chemical equilibrium.
We consider a situation that $\phi_{m}$ goes out of equilibrium 
at a time $t_{0}$. $X_{m}$ at a time $t_{0} + \Delta t$ 
($\Delta t$ is an infinitesimal time interval) is obtained by solving 
the Eq.~\eqref{Eq:Boltz_Xm} as 
\begin{equation}
\begin{split}
	X_{m} 
	= 
	X_{m}^{eq} + Ce^{-\langle \Gamma \rangle \Delta t /2}, 
	\label{Eq:Xm_1}
\end{split}     
\end{equation}
where $C$ is a constant. 
As long as the inequality 
\begin{equation}
\begin{split}
	\langle \Gamma \rangle_{m \leftrightarrow 
	\phi^{\dagger} \phi} \Delta t \gg 1
	\label{Eq:cond_1}
\end{split}     
\end{equation}
is satisfied, $m$ immediately goes back to equilibrium. Hence 
this inequality stands for the equilibrium condition of $m$. 

In the derivation of the condition~\eqref{Eq:cond_1}, we implicitly 
assume that $X_{m}^{eq}$ is constant in an interval $\Delta t$. 
It is necessary for justification for the condition~\eqref{Eq:cond_1} 
to guarantee the inequality $\Delta X_{m}^{eq}/X_{m}^{eq} 
\ll 1$ in $\Delta t$. $\Delta X_{m}^{eq}/X_{m}^{eq}$ in 
$\Delta t$ is computed as follows, 
\begin{equation}
\begin{split}
	\frac{\Delta X_{m}^{eq}}{X_{m}^{eq}} 
	&= 
	\frac{\Delta t}{X_{m}^{eq}}
	\left\{ 
	R^{3} \frac{\Delta n_{m}^{eq}}{\Delta t}
	+ n_{m}^{eq} \left( 3\dot{R} R^{2}  \right)
	\right\}
	\\&= 
	\Delta t 
	\left\{ 
	\frac{\Delta \text{log}n_{m}^{eq}}{\Delta t} + 3H
	\right\}. 
	\label{Eq:DelXeq/Xeq_1}
\end{split}     
\end{equation}
The first term for $T_{0} < m_{m}$, where 
$T_{0}$ represents the temperature which $m$ goes 
out of equilibrium, is 
\begin{equation}
\begin{split}
	\frac{\Delta \text{log}n_{m}^{eq}}{\Delta t} 
	&= 
	- H \left( \frac{3}{2} + \frac{m_{m}}{T} \right). 
	\label{Eq:DelLog/Delt}
\end{split}     
\end{equation}
Thus the inequality $\Delta X_{m}^{eq}/X_{m}^{eq} \ll 1$ is 
rewritten in terms of $m_{m}$ and $T$ as 
\begin{equation}
\begin{split}
	\frac{\Delta X_{m}^{eq}}{X_{m}^{eq}} 
	&= 
	\left( \frac{3}{2} - \frac{m_{m}}{T} \right) 
	\Delta t H
	\\&\simeq 
	- \frac{m_{m}}{T} \Delta t H. 
	\label{Eq:DelXeq/Xeq_2}
\end{split}     
\end{equation}
The approximation from the first line to the second line holds for 
$m_{m} \gg T$.

As a result, by combining the conditions \eqref{Eq:cond_1} 
and \eqref{Eq:DelXeq/Xeq_2}, we find the condition to maintain the 
equilibrium of $m$ and the SM fields as 
\begin{equation}
\begin{split}
	\frac{T}{m_{m}} 
	\frac{\langle \Gamma \rangle_{m 
	\leftrightarrow \phi^{\dagger} \phi}}{H} 
	\gg 1 . 
	\label{Eq:cond_result}
\end{split}     
\end{equation}
We can convert the condition in terms of the model parameters as 
follows 
\begin{equation}
\begin{split}
	&
	\frac{T}{m_{m}} 
	\frac{\langle \Gamma \rangle_{m 
	\leftrightarrow \phi^{\dagger} \phi}}{H} 
	\simeq 
	\\&   
	\left( \frac{g_{m\phi^\dagger\phi}}{10^{-7}} \right)^{2} 
	\left( \frac{106.75}{g_{*}} \right)^{1/2} 
	\left( \frac{100\,\text{GeV}}{T} \right)
	\gg 1. 
	\label{Eq:cond_tildec3}
\end{split}     
\end{equation}
Hence $m$ goes out of the equilibrium of $m$ 
and the SM fields for $g_{m\phi^\dagger\phi} \lesssim 10^{-7}$. The 
non-equilibration of $m$ indirectly gives rise to the 
$d$ decoupling from the SM thermal bath via the 
$d$-$m$ scattering. As a result the $m$ 
life-time controls the $d$ relic density. We will numerically 
check these results in next section. 

Note that there exist scattering processes with SM fermions 
$\psi_{i}$ or SM gauge bosons, e.g., 
$m \phi \leftrightarrow \psi_{i} \bar{\psi}_{i}$, 
$m \phi \leftrightarrow W^{+} W^{-}$, and so on, and 
they can contribute to the $m$ thermalization. These 
contributions are however negligible. This is understood as follows. 
As is noted above, key ingredient in our scenario is that $m$ 
goes out of the equilibrium between $m$ and SM fields, 
which leads non-familiar relic density of dark matter. $m$ 
goes out of the equilibrium for $g_{m\phi^\dagger\phi} \lesssim 10^{-7}$. 
As long as $g_{dd\phi^\dagger\phi} \gtrsim 10^{-4}$ which is minimum value for the 
$m$ thermalization, the reaction $mm \leftrightarrow \phi \phi$ dominates over the 
scatterings processes with SM fermions or SM gauge bosons. 
Hence we can omit the scattering processes with SM particles 
via $g_{m\phi^\dagger\phi}$ coupling.

\section{Numerical Results}
\label{sec:numerical-results}

We here present our results on the numerical computations for 
the DM relic density. 

The evolutions of $d$ and $m$ particles are illustrated
in Sec.~\ref{Sec:evo},\ref{Sec:evo2} and Sec.~\ref{Sec:evo3}
with numerical results computed at several reference points. 
We assume $m_d=m_m$ in these reference points, since the DM 
evolution behaves quite differently than the scenario with $m_d \gg m_m$.
Motivated by the unified model of the DM and the mediator,
we consider 
$g_{dd\phi^\dagger\phi} = 
g_{mm\phi^\dagger\phi}$ 
case throughout in this section.
We emphasize that all of the results shown below 
is not affected, however, 
even if $g_{dd\phi^\dagger\phi} = 0$
as long as $g_{ddmm}\ne 0$, $g_{mm\phi^\dagger\phi}\ne 0$.

The parameter dependences of the DM relic density are shown in
Sec.~\ref{Sec:lambda} and Sec.~\ref{Sec:life} assuming
$m_d=m_m$.
We find that it is necessary to take into account the 
non-unity ratio $n_{m}/n_{m}^{eq} \neq 1$ in the Boltzmann 
equations~\eqref{Eq:Boltz_1_1} and \eqref{Eq:Boltz_1_2}. 
The important role of $n_{m}/n_{m}^{eq} \neq 1$ should be emphasized.

In Sec.~\ref{Sec:mm/md}, we plot the parameter regions 
consistent with the observed DM relic density in the 
$r$-$g_{ddmm}$ plane.   Here the mass ratio $m_m/m_d$
is denoted by $r$.
We find numerically that, if the life-time of the mediator
particle $m$ is longer than the DM decoupling time from the 
mediator, the mass ratio $r$ cannot exceed $\sim 0.95$
in order to account for the observed DM relic abundance.
On the other hand, if we introduce a life-time of $m$ 
comparable with or shorter than the DM decoupling time, 
the situation changes drastically.
Our mechanism yields the observed 
DM relic density in the secluded scenario even for a completely
degenerated $d$-$m$ system, with which
we deduce the upper bound on the mediator life-time.

The validity of the Boltzmann equation Eq.(\ref{Eq:Boltz_2})
in which both $d$, $m$ and Higgs are assumed to feel the same 
temperature is studied in Sec.~\ref{Sec:T}.

\subsection{Evolution example 1} 
\label{Sec:evo}

\begin{figure}[h!]
\begin{center}
\includegraphics[clip, width=87mm]{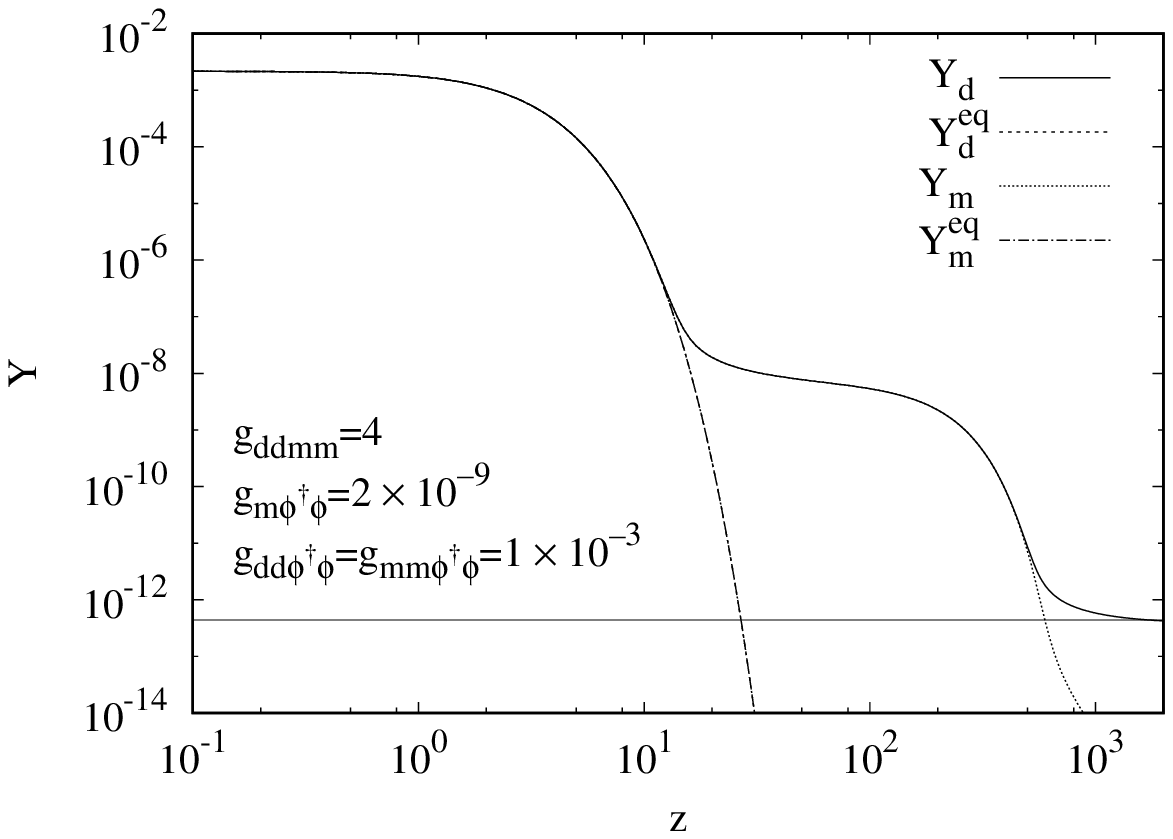}
\end{center}
\caption{Evolutions of $d$ and $m$. 
The observed DM relic density $Y_{d}^{\rm obs}=(4.330 \pm 0.036)\times 10^{-13}$ 
is shown by the horizontal band.
$m_{d} = m_{m}= 1\,\text{TeV}$,
$g_{ddmm} = 4$, 
$g_{m\phi^\dagger\phi} = 2 \times 10^{-9}$, and 
$g_{dd\phi^\dagger\phi} = g_{mm\phi^\dagger\phi} = 1 \times 10^{-3}$. 
}
\label{Fig:evolution_1}
\begin{center}
\includegraphics[clip, width=87mm]{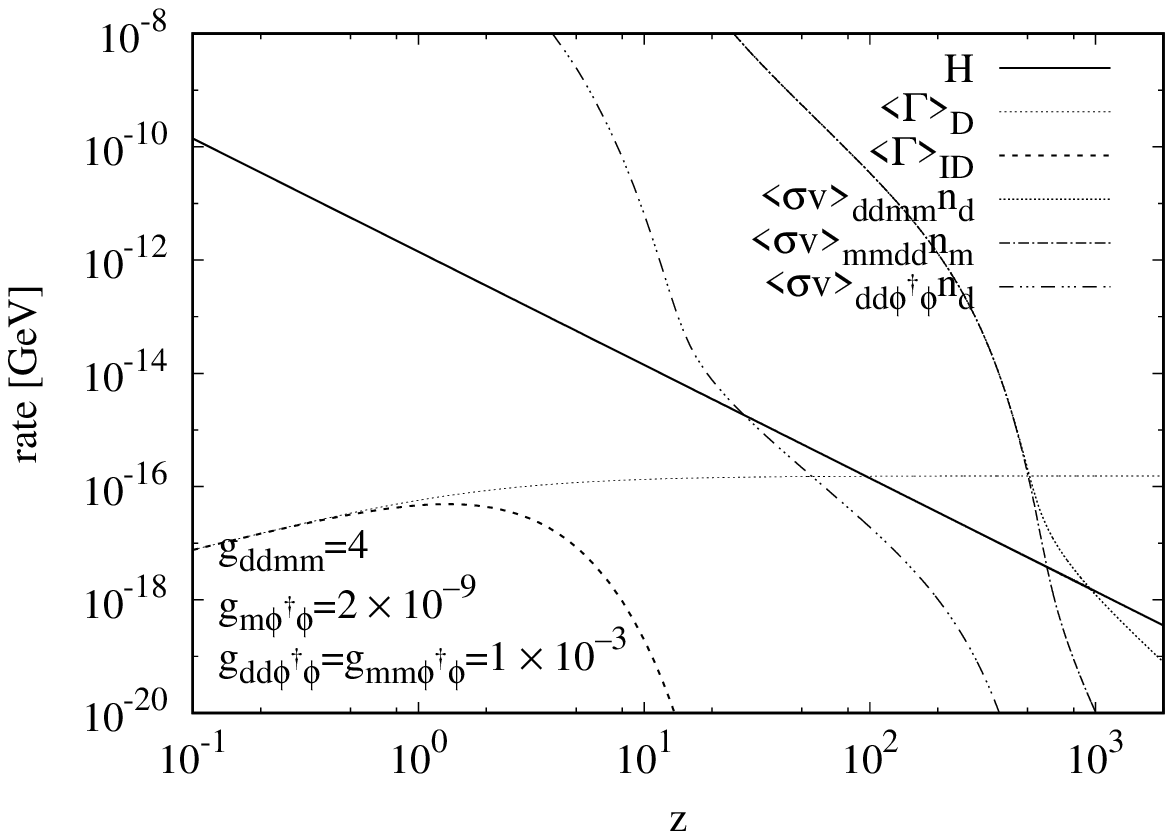}
\end{center}
\caption{The Hubble rate $H$ and 
interaction rates, $\langle \Gamma \rangle_\text{D}$, 
$\langle \Gamma \rangle_\text{ID}$, 
$\langle \sigma v \rangle_{ddmm} n_{d}$, 
$\langle \sigma v \rangle_{mmdd} n_{m}$, 
and $\langle \sigma v \rangle_{dd\phi\dagger\phi} n_{d}$.
$m_{d} = m_{m}= 1\,\text{TeV}$,
$g_{ddmm} = 4$, 
$g_{m\phi^\dagger\phi} = 2 \times 10^{-9}$, and 
$g_{dd\phi^\dagger\phi} = g_{mm\phi^\dagger\phi} = 1 \times 10^{-3}$. 
}
\label{Fig:comp_rate}
\end{figure}

We here give an example of the number density evolutions 
for the DM and mediator particles.
A typical evolution of $Y_d=n_d/s$ ($Y_m=n_m/s$) 
is shown in  Fig.~\ref{Fig:evolution_1}, with
$Y_d$ ($Y_m$) being the number density normalized by the 
entropy density $s$ for $d$ ($m$).
The horizontal axis shows $z = m_{d}/T$.
The mediator $m$ and the DM $d$ are assumed to degenerate in mass 
$m_{d} = m_{m}= 1\,\text{TeV}$.
We take  
$g_{ddmm} = 4$, 
$g_{m\phi^\dagger\phi} = 2 \times 10^{-9}$, and 
$g_{dd\phi^\dagger\phi} = g_{mm\phi^\dagger\phi} = 1 \times 10^{-3}$ 
as a reference point in this plot. 
The Hubble rate $H$, 
thermal averaged decay rate and inverse decay rate 
of $m \leftrightarrow \phi^{\dagger} \phi$ 
($\langle \Gamma \rangle_\text{D}$ and 
$\langle \Gamma \rangle_\text{ID}$),  
and interaction rates of 
$dd \to mm$, 
$mm \to dd$, 
and 
$dd \to \phi^{\dagger} \phi$ 
are shown in Fig.~\ref{Fig:comp_rate}
with the same parameter set. 
As we see in Fig.~\ref{Fig:comp_rate}, there exist 
characteristic time scales which play important roles in the determination 
of the evolutions.
They are
\begin{itemize}
\item[(a)] The time scale $z_{dd\phi^\dagger\phi}$ at which
the both DM and mediator go out of chemical equilibrium 
from the SM thermal bath.
We find it is $z_{dd\phi^\dagger\phi}\simeq 27$ in Fig.~\ref{Fig:comp_rate}.
This scale can be determined by the condition
\begin{equation}
 \VEV{\sigma v}_{dd\phi^\dagger\phi} n_d =H. 
\end{equation}
\item[(b)] The scale $z_{m\phi^\dagger\phi}\simeq 96$ determined by the mediator 
life-time,
\begin{equation}
  \VEV{\Gamma}_{\rm D}=H .
\end{equation}
The mediator decay starts to affect the evolution of the system
after $z>z_{m\phi^\dagger\phi}$.

\item[(c)] The time scale $z_{\rm E}$ until when the detailed balance
between the $dd \to mm$ and $mm \to dd$ processes is held.
The $dd \leftrightarrow mm$ detailed balance ends at $z_E$.
Due to the rapid decreasing of the mediator density through the mediator decay,
the mediator density $Y_m$ falls below $Y_d$ after $z_{\rm E}$.
The mediator decay rate
$\VEV{\Gamma}_{\rm D}$ exceeds the interaction rate of
$dd \leftrightarrow mm$ for $z>z_{\rm E}$.
We see $z_{\rm E}\simeq 510$ with the present set of parameters.

\item[(d)] The time scale $z_{ddmm}$ when the $dd\to mm$ interaction
rate becomes slower than the Hubble rate.
The DM density is frozen to its final abundance after $z_{ddmm}$, 
decoupled from the mediator particle $m$.
We find $z_{ddmm} \simeq 950$ in Fig.~\ref{Fig:comp_rate}.
\end{itemize}

The DM evolution in this setup exhibits distinctive ``terrace'' structure
as shown in Fig.~\ref{Fig:evolution_1},
which can be understood 
step by step in terms of these characteristic time scales.

In the beginning of the evolution 
 ($z \ll z_{dd\phi^\dagger\phi}\simeq 27$), our scenario 
traces familiar 
DM evolution in the Higgs portal DM scenario.
The DM $d$ is thermalized through the 
process $dd \leftrightarrow \phi^{\dagger} \phi$ whose rate is much larger 
than $H$. 
The equilibrium between $\phi_{m}$ and the background SM fields is 
also achieved through the process $mm \leftrightarrow \phi^{\dagger} \phi$. 

At the next stage ($z_{dd\phi^\dagger\phi} \lesssim z \lesssim z_{m\phi^\dagger\phi}\simeq 96$), 
the DM density $Y_d$ is  temporarily ``frozen'' to a value ($Y_d \sim 10^{-8}$) 
much higher than the corresponding value ($Y_d \sim 4\times 10^{-13}$) in the 
well-known Higgs portal DM scenario.
We call this phenomenon ``fake'' freeze-out of the DM.
The larger value of $Y_d \sim 10^{-8}$ is 
because the DM interacts with 
the SM much weaker 
than the Higgs portal DM in the present model
and thus it decouples from the SM at an earlier time.
Note that the mediator $m$ goes out of the equilibrium with the SM
simultaneously with the DM ``fake'' freeze-out.
This situation holds even in the $g_{dd\phi^\dagger\phi} \ll g_{mm\phi^\dagger\phi}$ case
due to the sizable interactions among $m$ and $d$. 

The mediator decay becomes active after $z_{m\phi^\dagger\phi} \simeq 96$.
The inverse decay rate $\VEV{\Gamma}_{\rm ID}$, on the other hand, is negligibly
small.
The $m$ density in a comoving volume starts to decrease exponentially 
after $z\simeq z_{m\phi^\dagger\phi}$ through the mediator 
decay $m \to \phi^{\dagger} \phi$.
The temporarily ``frozen'' mediator density $Y_m$ is 
then thawed by the mediator 
decay.
We see in Fig.~\ref{Fig:evolution_1}, 
due to the strong interaction $dd\leftrightarrow mm$,
the DM density $Y_d$ tracks $Y_m$ until $z\simeq z_{\rm E}\simeq 510$.
We thus find the DM density $Y_d$ decreases drastically after its
decoupling from the SM thermal bath.
It is important to emphasize that the decoupling of $d$ and 
$m$ from the SM thermal bath does not imply 
the freeze-out of their densities. 

Once the $m$ decay rate exceeds the interaction rate of 
$dd \leftrightarrow mm$, 
the rapidly-decreasing $m$ density breaks the detailed 
balance between the interactions 
$mm \to dd$ 
and 
$dd \to mm$. 
Note that the DM density is still decreasing if 
the interaction rate of $dd \to mm$ 
is larger than the Hubble rate $H$.
The DM density is fixed
to its final abundance only at the last stage ($z \gtrsim z_{ddmm}\simeq 950$)
when  
the interaction rate of $dd \to mm$ becomes smaller than $H$ .

\subsection{Evolution  example 2}
\label{Sec:evo2}

\begin{figure}[h!]
\begin{center}
\includegraphics[clip, width=87mm]{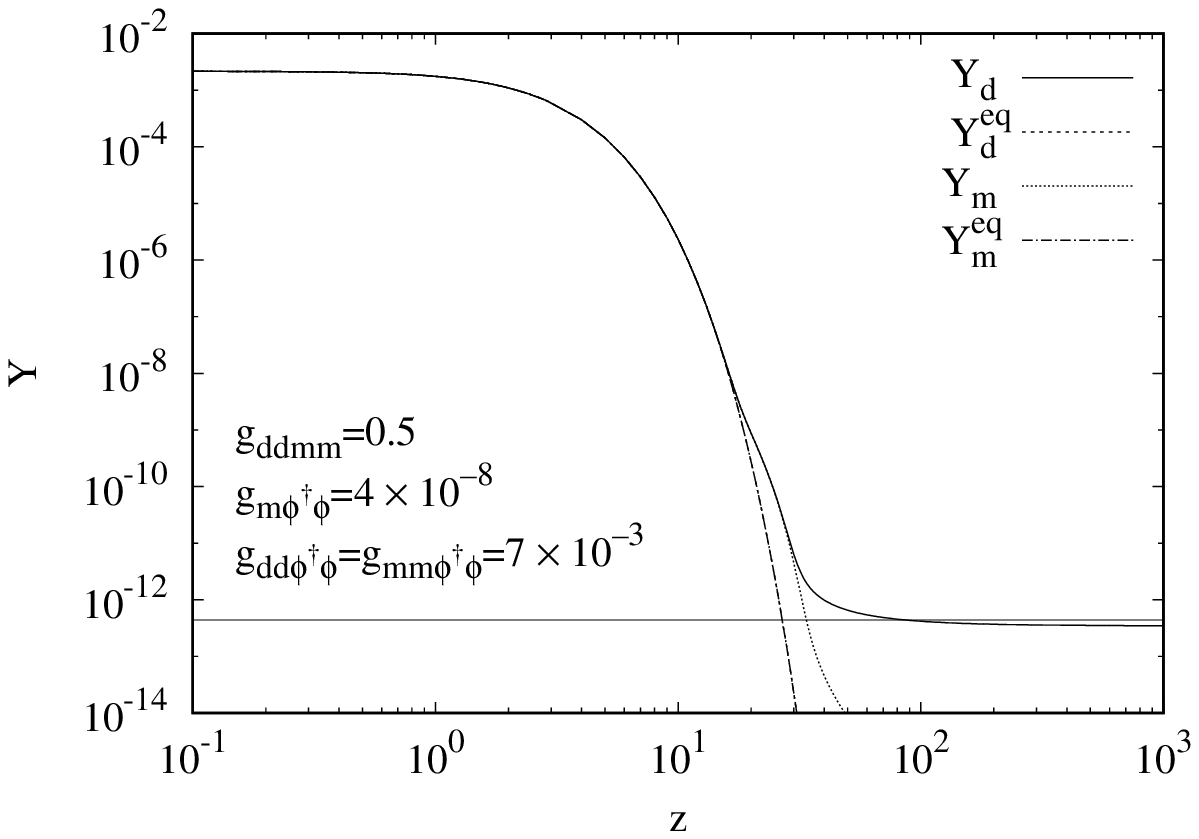}
\end{center}
\caption{Evolutions of $d$ and $m$. 
The observed DM  relic density $Y_{d}^{\rm obs}=(4.330 \pm 0.036)\times 10^{-13}$ 
is shown by the horizontal band.
$m_{d} = m_{m}= 1\,\text{TeV}$,
$g_{ddmm} = 0.5$, 
$g_{m\phi^\dagger\phi} = 4 \times 10^{-8}$, and 
$g_{dd\phi^\dagger\phi} = g_{mm\phi^\dagger\phi}= 7 \times 10^{-3}$. 
}
\label{Fig:evolution_2}
\begin{center}
\includegraphics[clip, width=87mm]{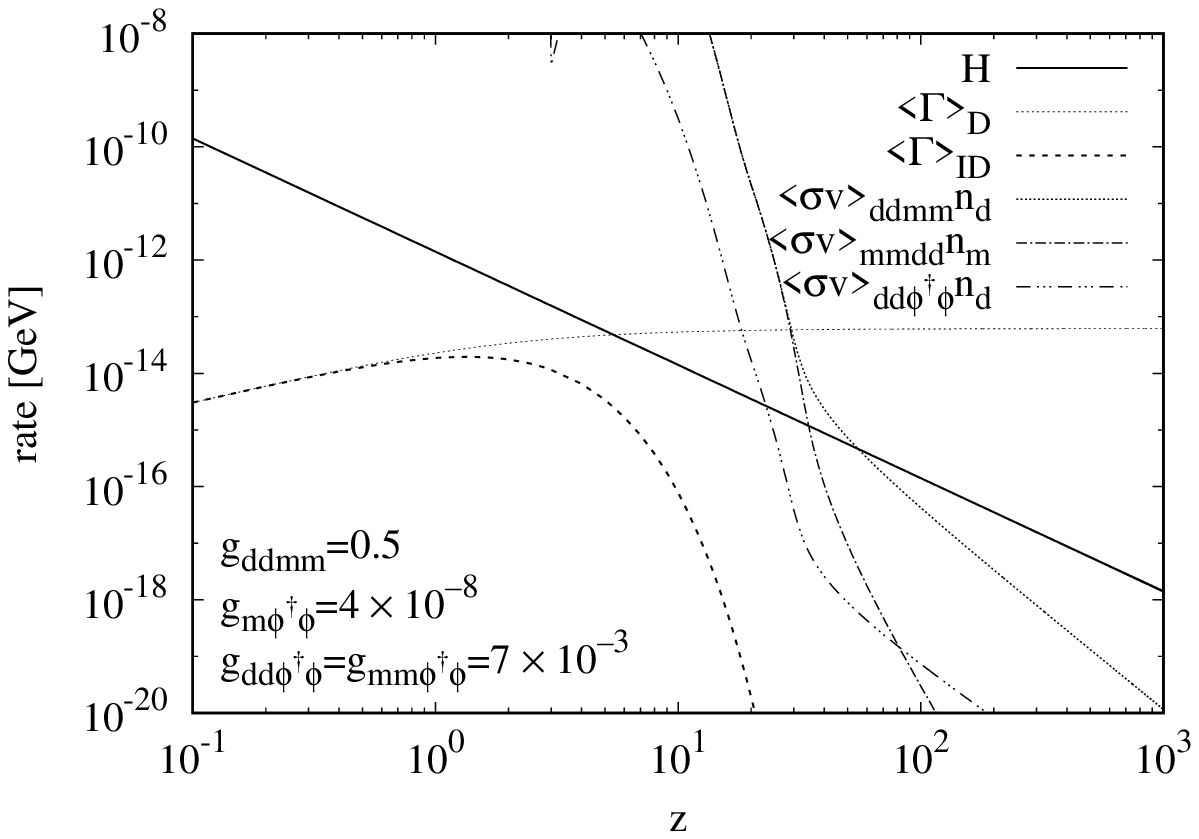}
\end{center}
\caption{The Hubble rate $H$ and 
interaction rates, $\langle \Gamma \rangle_\text{D}$, 
$\langle \Gamma \rangle_\text{ID}$, 
$\langle \sigma v \rangle_{ddmm} n_{d}$, 
$\langle \sigma v \rangle_{mmdd} n_{m}$, 
and $\langle \sigma v \rangle_{dd\phi^\dagger\phi} n_{d}$.
$m_{d} = m_{m}= 1\,\text{TeV}$,
$g_{ddmm} = 0.5$, 
$g_{m\phi^\dagger\phi} = 4 \times 10^{-8}$, and 
$g_{dd\phi^\dagger\phi} = g_{mm\phi^\dagger\phi}= 7 \times 10^{-3}$. 
}
\label{Fig:comp_rate_2}
\end{figure}

Another example of typical evolution is shown in Fig.~\ref{Fig:evolution_2},
which differs qualitatively from the example we had shown in the previous
subsection.
In this plot, we take parameters 
$m_d=m_m=1$TeV, $g_{ddmm}=0.5$, $g_{m\phi^\dagger\phi}=4 \times 10^{-8}$, and 
$g_{dd\phi^\dagger\phi}=g_{mm\phi^\dagger\phi}=7\times 10^{-3}$,
which lead to
$z_{dd\phi^\dagger\phi} \simeq 23$ , 
$z_{m\phi^\dagger\phi} \simeq 6$, 
$z_{\rm E} \simeq 29$,
and
$z_{ddmm} \simeq 56$ as shown in Fig.~\ref{Fig:comp_rate_2}.
We therefore find $z_{dd\phi^\dagger\phi}>z_{m\phi^\dagger\phi}$, in contrast to 
the inequality $z_{dd\phi^\dagger\phi}<z_{m\phi^\dagger\phi}$ we had in 
the previous subsection.
As a result, the distinctive ``terrace'' structure we observed in the
previous subsection disappears in Fig.~\ref{Fig:evolution_2}.
Instead, we see a change of slope at the scale $z_{dd\phi^\dagger\phi}$ in the plot of $Y_d$.

\subsection{Evolution  example 3}
\label{Sec:evo3}

\begin{figure}[h!]
\begin{center}
\includegraphics[clip, width=87mm]{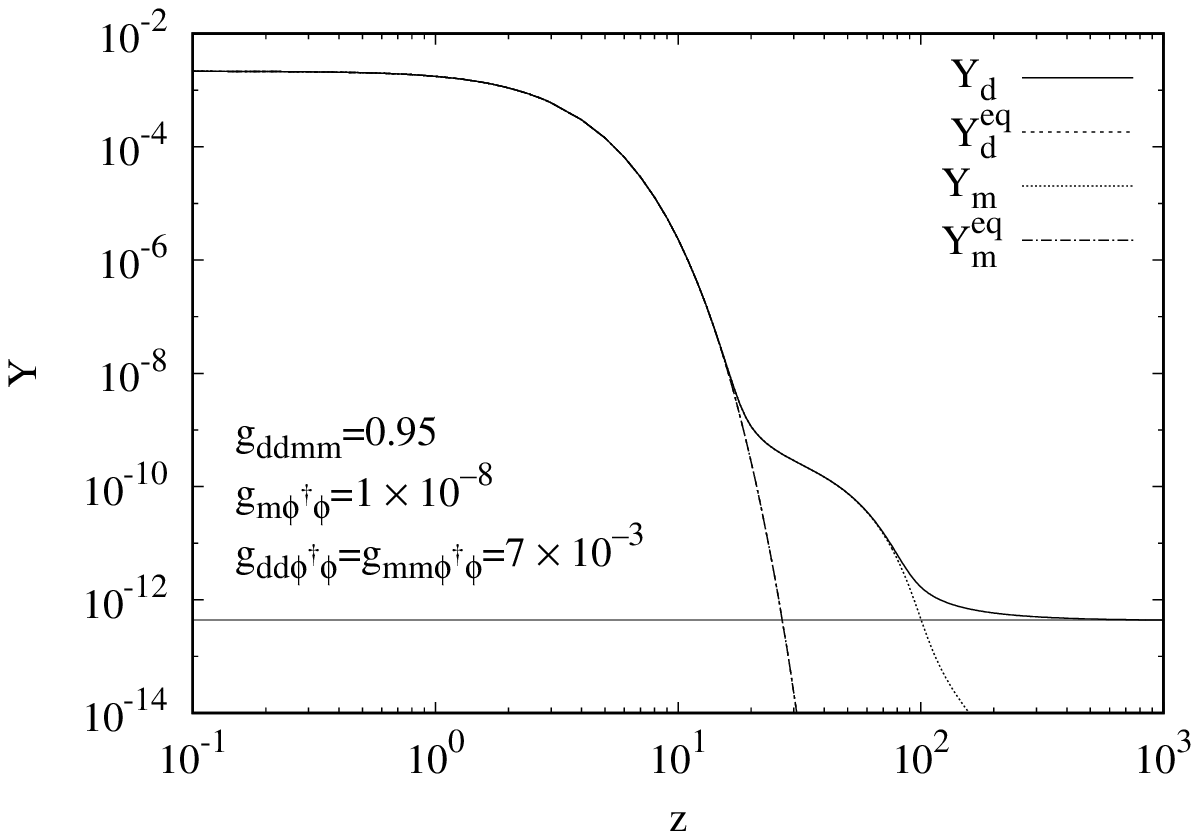}
\end{center}
\caption{Evolutions of $d$ and $m$. 
The observed DM relic density $Y_{d}^{\rm obs}=(4.330 \pm 0.036)\times 10^{-13}$ 
is shown by the horizontal band.
$m_{d} = m_{m}= 1\,\text{TeV}$,
$g_{ddmm} = 0.95$, 
$g_{m\phi^\dagger\phi} = 1 \times 10^{-8}$, and 
$g_{dd\phi^\dagger\phi} = g_{mm\phi^\dagger\phi} = 7 \times 10^{-3}$. 
}
\label{Fig:evolution_3}
%%%
%%%
\begin{center}
\includegraphics[clip, width=87mm]{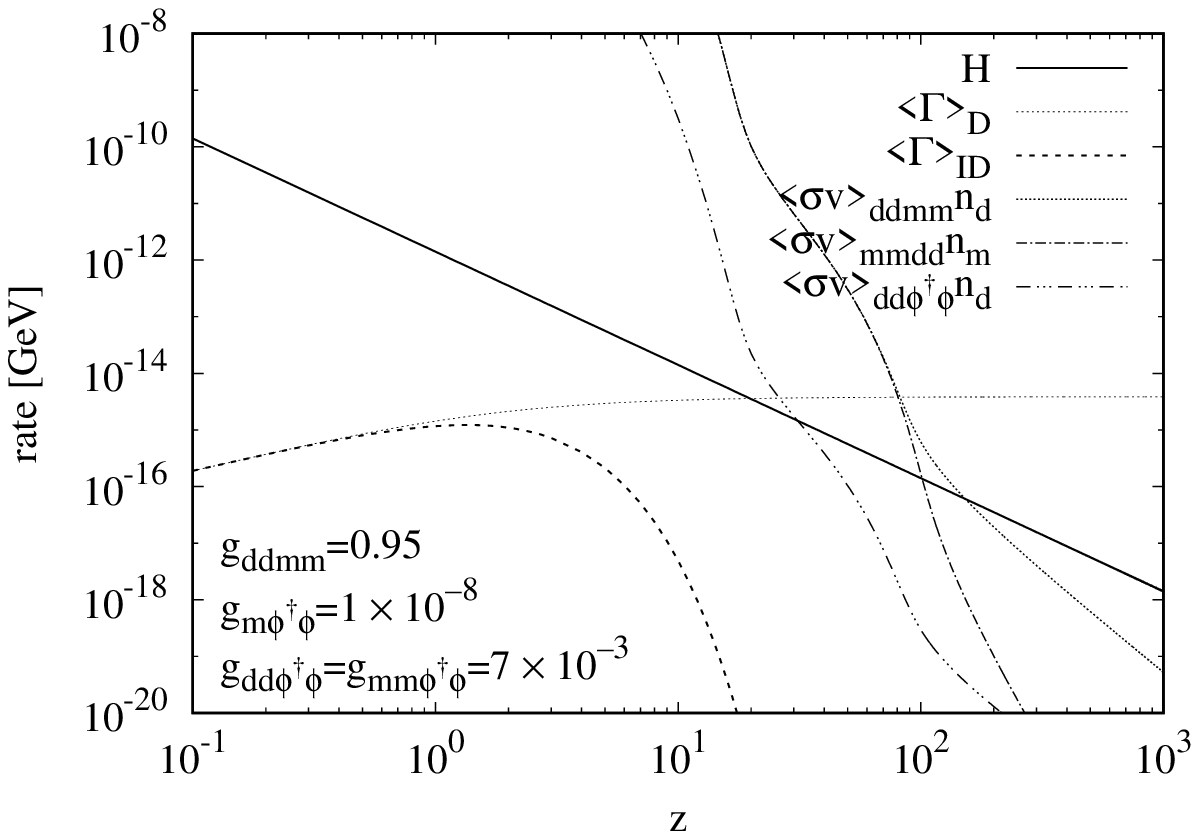}
\end{center}
\caption{The Hubble rate $H$ and 
interaction rates, $\langle \Gamma \rangle_\text{D}$, 
$\langle \Gamma \rangle_\text{ID}$, 
$\langle \sigma v \rangle_{ddmm} n_{d}$, 
$\langle \sigma v \rangle_{mmdd} n_{m}$, 
and $\langle \sigma v \rangle_{dd\phi^\dagger\phi} n_{d}$.
$m_{d} = m_{m}= 1\,\text{TeV}$,
$g_{ddmm} = 0.95$, 
$g_{m\phi^\dagger\phi} = 1 \times 10^{-8}$, and 
$g_{dd\phi^\dagger\phi} = g_{mm\phi^\dagger\phi} = 7 \times 10^{-3}$. 
}
\label{Fig:comp_rate_3}
\end{figure}

Here we give an example in which the mediator decay time scale $z_{m\phi^\dagger\phi}$ 
coincides approximately with the DM decoupling scale $z_{dd\phi^\dagger\phi}$.
This situation happens with parameters
$m_d=m_m=1$TeV, $g_{ddmm}=0.95$, $g_{m\phi^\dagger\phi}=1 \times 10^{-8}$, and 
$g_{dd\phi^\dagger\phi}=g_{mm\phi^\dagger\phi}=7\times 10^{-3}$.
Corresponding time scales are
$z_{dd\phi^\dagger\phi} \simeq 31$ , 
$z_{m\phi^\dagger\phi} \simeq 20$, 
$z_{\rm E} \simeq 82$,
and
$z_{ddmm} \simeq 152$.
See Fig.~\ref{Fig:evolution_3} and Fig.~\ref{Fig:comp_rate_3} for the
behavior of evolution and interaction rates, respectively.
The mediator decay affects the evolution immediately after the DM decoupling
from the SM thermal bath.  We see no clear terrace structure nor simple slope-change 
in the DM evolution shown in Fig.~\ref{Fig:evolution_3}.

%%%%%%%%%%%%%%%%%%%%%%%%%%%%%%%%%%%%%%%%
\subsection{$g_{ddmm}$ dependence} \label{Sec:lambda}   %%%%%%%%%%%%
%%%%%%%%%%%%%%%%%%%%%%%%%%%%%%%%%%%%%%%%

\begin{figure}[h!]
\begin{center}
\includegraphics[clip, width=87mm]{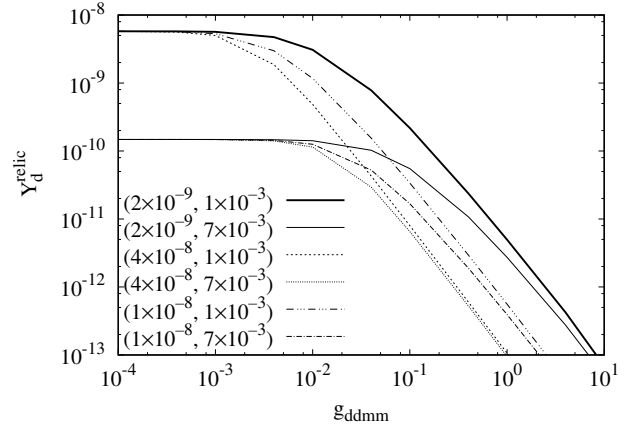}
\end{center}
\caption{$g_{ddmm}$ dependence of  $d$ relic density 
for each parameter set. Values in the legend represent the 
parameter set $(g_{m\phi^\dagger\phi}, g_{dd\phi^\dagger\phi} = g_{mm\phi^\dagger\phi})$. 
}
\label{Fig:LamDep}
\end{figure}

\begin{figure}[h!]
\begin{center}
\includegraphics[clip, width=87mm]{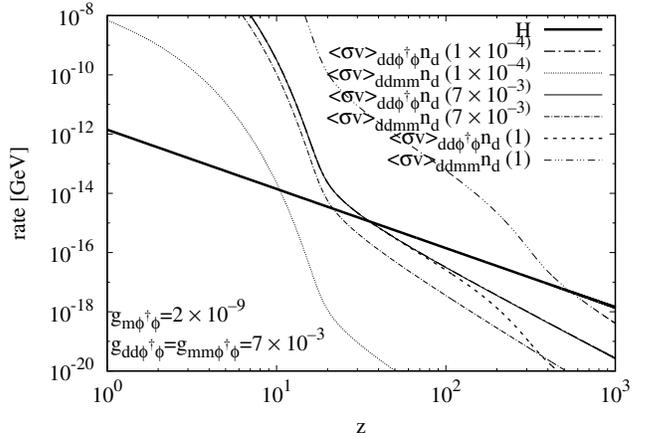}
\end{center}
\caption{Comparison of Hubble rate $H$ and interaction rates 
of $dd \to \phi^{\dagger} \phi$ and 
$dd \to mm$. Numbers in the 
legend represent $g_{ddmm}$ values.}
\label{Fig:comp}
\end{figure}

We study the $g_{ddmm}$ dependence of the $d$ relic density. 
Fig.~\ref{Fig:LamDep} shows the DM relic density as a 
function of $g_{ddmm}$. 
We take $m_{d} = m_{m} = 1\,\text{TeV}$. 
Parameters taken for the analysis are 
$(g_{m\phi^\dagger\phi}, g_{dd\phi^\dagger\phi} = g_{mm\phi^\dagger\phi}) 
= 
(2 \times 10^{-9}, 1 \times 10^{-3})$, 
$(2 \times 10^{-9}, 7 \times 10^{-3})$, 
$(4 \times 10^{-8}, 1 \times 10^{-3})$, 
$(4 \times 10^{-8}, 7 \times 10^{-3})$, 
$(1 \times 10^{-8}, 1 \times 10^{-3})$, 
and 
$(1 \times 10^{-8}, 7 \times 10^{-3})$, 
respectively. 

Larger $g_{ddmm}$ provides smaller $d$ relic density. 
This is because longer equilibrium between $d$ and 
$m$ can be achieved by larger $g_{ddmm}$, which delays 
the decoupling between $d$ and $m$. 
As we show in Appendix~\ref{Sec:ana_est}, 
the semi-analytic formula~\eqref{eq:Boltz_5c} for the DM relic abundance 
$Y_d^{\rm relic}$ is actually inversely proportional to $g_{ddmm}^2$ 
and supports this understanding. 

On the other hand, smaller $g_{ddmm}$ gives larger relic density.
For each value of $g_{dd\phi^\dagger\phi}$, 
the relic density approaches to its asymptotic value in the 
$d$-$m$ collision-less limit, which corresponds 
to the DM relic density controlled by the $g_{dd\phi^\dagger\phi}$ 
in the Higgs portal scenario.
The DM relic density becomes almost insensitive to the value
of $g_{ddmm}$ for $g_{ddmm} \lesssim g_{dd\phi^\dagger\phi}$.
This is understood as follows. 
Fig.~\ref{Fig:comp} shows the interaction rates of $dd\to mm$ and 
$dd \to \phi^\dagger\phi$ for 
$g_{ddmm}=10^{-4}$, $7\times 10^{-3}$ and $1$.
Other coupling strengths are taken as
$g_{dd\phi^\dagger\phi}=g_{mm\phi^\dagger\phi}=7\times 10^{-3}$
and $g_{m\phi^\dagger\phi}=2\times 10^{-9}$.
We see the DM-mediator decoupling epoch ($\VEV{\sigma v}_{ddmm} n_d =H$)
is earlier than the DM-Higgs decoupling 
($\VEV{\sigma v}_{dd\phi^\dagger\phi} n_d = H$)
for $g_{ddmm}<g_{dd\phi^\dagger\phi}$.
There is therefore no $g_{ddmm}$ dependence of the relic density 
for $g_{ddmm} \lesssim g_{dd\phi^\dagger\phi}$. 

Hence, our mechanism to reduce the DM relic density 
works only when the $g_{ddmm}$ coupling is stronger 
than the DM-Higgs coupling $g_{dd\phi^\dagger\phi}$ 
in the Higgs portal DM scenario.

%%%%%%%%%%%%%%%%%%%%%%%%%%%%%%%%%%%%%%%%
\subsection{Mediator life-time dependence} \label{Sec:life} %%%%%%%%%%
%%%%%%%%%%%%%%%%%%%%%%%%%%%%%%%%%%%%%%%%

\begin{figure}[h!]
\begin{center}
\includegraphics[clip, width=87mm]{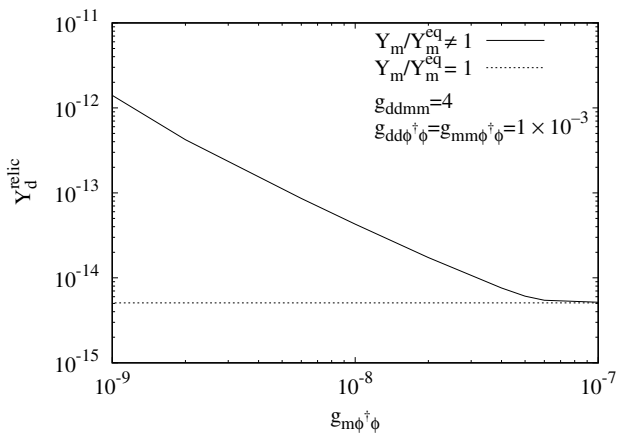}
\end{center}
\caption{$g_{m\phi^\dagger\phi}$ dependence of the DM relic density 
for 
$Y_{m}/Y_{m}^{eq} \neq 1$ (solid line)
and 
$Y_{m}/Y_{m}^{eq} = 1$ (dotted line). }
\label{Fig:c3Dep}
\end{figure}

\begin{figure}[h!]
\begin{center}
\includegraphics[clip, width=87mm]{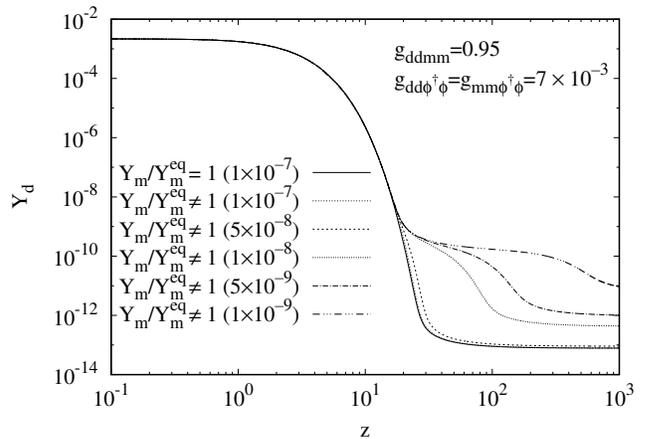}
\end{center}
\caption{Comparison of $d$ evolution for 
$Y_{m}/Y_{m}^{eq} = 1$ (solid line) and 
$Y_{m}/Y_{m}^{eq} \neq 1$ (other lines). 
Numbers in the legend correspond to $g_{m\phi^\dagger\phi}$ 
in each calculation.}
\label{Fig:EvoComparizon_Y/Yeq}
\end{figure}

If the mediator were in chemical equilibrium with the
thermal bath, $Y_{m}/Y_{m}^{eq}=1$ and thus $n_m/n_m^{eq}=1$.
The DM Boltzmann equation (\ref{Eq:Boltz_1_1}) would then 
be separated from the mediator Boltzmann equation (\ref{Eq:Boltz_1_2}) 
completely.
The DM relic abundance would therefore be insensitive 
to the mediator properties such as the mediator life-time
($g_{m\phi^\dagger\phi}$ coupling).
In the reality, however, the mediator departs from its
chemical equilibrium almost simultaneously with 
the ``fake'' freeze-out epoch of the DM.  We need to
take into account effects of $Y_{m}/Y_{m}^{eq} \ne 1$
in our computations of the DM relic density. 
Solving the coupled Boltzmann equations (\ref{Eq:Boltz_1_1}) 
and (\ref{Eq:Boltz_1_2}), we obtain the DM relic density as 
plotted in the solid line in Fig.~\ref{Fig:c3Dep} as a function 
of $g_{m\phi^\dagger\phi}$. 
On the other hand, if the chemical equilibrium
of the mediator particle $n_m=n_m^{eq}$ were satisfied 
in (\ref{Eq:Boltz_1_1}), 
we would obtain the dotted line result in Fig.~\ref{Fig:c3Dep}.
We see the coupling $g_{m\phi^\dagger\phi}$ controls the
DM relic density. 
The effects of the departure of the mediator chemical equilibrium 
$Y_{m}/Y_{m}^{eq} \ne 1$ are significant.

The relic density decreases with decreasing life-time 
(increasing $g_{m\phi^\dagger\phi}$).
It approaches to the value of $Y_{m}/Y_{m}^{eq} = 1$ around 
$g_{m\phi^\dagger\phi} \simeq 10^{-7}$. The behavior of the relic 
density is understood as follows.
For $g_{m\phi^\dagger\phi} \gtrsim 10^{-7}$ as derived in 
Eq.~\eqref{Eq:cond_tildec3},
the mediator keeps its chemical equilibrium with the SM Higgs. 
The $m$ density in (quasi-)equilibrium exponentially drops, 
and guides the over-abundant $d$ density to the observed one. 

Fig.~\ref{Fig:EvoComparizon_Y/Yeq} shows the evolutions of 
$Y_{d}$ for each $g_{m\phi^\dagger\phi}$. 
Too long $m$ life-time keeps over-densities of $m$ 
for a long period, and leads to a mild damping of $d$. 
Freeze-out of $Y_{d}$ occurs at large $z$ due to a large deviation 
of $Y_{m}/Y_{m}^{eq}$ from unity, and hence $Y_{d}$ remains 
over-abundant.
On the other hand, a deviation of $Y_{m}/Y_{m}^{eq}$ from unity 
becomes smaller for shorter $m$ life-time. The Boltzmann 
equations~\eqref{Eq:Boltz_1_1} and \eqref{Eq:Boltz_1_2} and 
the relic density approach to those in familiar thermal relic scenarios. 

We here note the non-unities $Y_m/Y_m^{eq} \ne 1$ 
and $Y_d/Y_d^{eq} \ne 1$ also affect the evolutions of the $d$ and $m$ 
densities after the epoch $z_{E}$, when the detailed balance 
of the $dd \leftrightarrow mm$ process is broken.

%%%%%%%%%%%%%%%%%%%%%%%%%%%%%%%%%%%%%%%%
\subsection{
Parameter regions consistent with the observed DM relic abundance
} 
\label{Sec:mm/md}%%%%%%%
%%%%%%%%%%%%%%%%%%%%%%%%%%%%%%%%%%%%%%%%

\begin{figure}[h!]
\begin{center}
\includegraphics[clip, width=87mm]{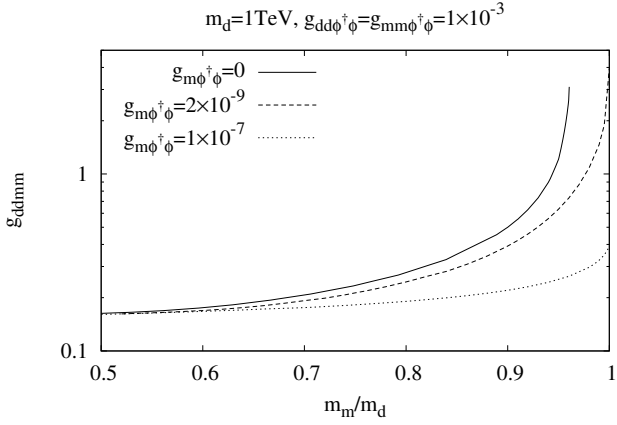}
\end{center}
\caption{
Parameters reproducing the central value of the observed DM relic
density $Y_d^{\rm obs}=4.330\times 10^{-13}$.
We take $m_d=1$TeV and 
$g_{dd\phi^\dagger\phi}=g_{mm\phi^\dagger\phi}=1 \times 10^{-3}$.
The solid, dashed and dotted lines correspond to the
cases with negligibly small $g_{m\phi^\dagger\phi}$,
$g_{m\phi^\dagger\phi}=2 \times 10^{-9}$, and 
$g_{m\phi^\dagger\phi}=1 \times 10^{-7}$, respectively.
}
\label{Fig:lambda_mm/md}
\end{figure}

We show parameters that can account 
for the central value of the observed DM relic density 
$Y_{d}^{\rm obs}=4.330\times 10^{-13}$
in Fig.~\ref{Fig:lambda_mm/md}.
We take $m_d = 1$ TeV and 
$g_{dd\phi^\dagger\phi} = 1 \times 10^{-3}$. 
The solid, dashed and dotted lines correspond to 
the case of negligibly small $g_{m\phi^\dagger\phi}$ (very late-time
decaying mediator), $g_{m\phi^\dagger\phi} =2 \times 10^{-9}$ and 
$1 \times 10^{-7}$, respectively. 
We see in this plot that for $r \equiv m_m/m_d \lesssim 1/2$ the 
coupling $g_{ddmm}$
required for the observed relic density gets close to an asymptotic 
value $g_{ddmm}\simeq 0.16$, almost independently of $g_{m\phi^\dagger\phi}$. 
This illustrates the fact that in secluded scenarios with a light mediator ($r\ll1$), 
the DM relic density is controlled only by $g_{ddmm}$ and the DM mass $m_d$, 
and becomes insensitive to $g_{m\phi^\dagger\phi}$ and $g_{dd\phi^\dagger\phi}$. 

On the other hand, as $r$ gets larger, the required coupling $g_{ddmm}$ 
also  becomes larger.
For the case of the negligibly small $g_{m\phi^\dagger\phi}$ 
(the very late time decaying mediator),  
especially, the coupling $g_{ddmm}$ goes over the unitarity bound
around $r \gtrsim 0.95$.
We thus find the upper bound on $r$ so as to explain the
observed DM relic density.
Thus, with the extremely long-lived mediator which survives even after
the DM decoupling from the mediator, 
the completely degenerated mediator setup ($r=1$) is not feasible 
to account for the observed DM relic density. 
This property is consistent with the numerical results for
the DM relic abundance done in the context of a right-handed
sneutrino-neutrino DM-mediator model\cite{Bandyopadhyay:2011qm}.

The situation changes drastically
if we consider a shorter life-time mediator. 
The secluded DM scenario with $r=1$ becomes viable
if the mediator life-time is comparable with or 
shorter than the DM decoupling time from the mediator. 
Actually, when $g_{m\phi^\dagger\phi}=2\times 10^{-9}$,
the coupling $g_{ddmm}\simeq 4$ is required at $r=1$ 
and is marginally consistent with the unitarity.
For the mediator with shorter life-time (larger $g_{m\phi^\dagger\phi}$), 
it is easier to 
find the parameter regions consistent with the observed
relic density and also with the unitarity.
We numerically obtain the lower limit on $g_{m\phi^\dagger\phi}$, 
$g_{m\phi^\dagger\phi} \gtrsim 2\times 10^{-9}$ for
$m_d=1$ TeV and $g_{dd\phi^\dagger\phi}=1\times10^{-3}$.

The required $g_{ddmm}$ coupling for $r=1$ decreases 
with increasing $g_{m\phi^\dagger\phi}$.
It approaches to an asymptotic value $g_{ddmm} \simeq 0.4$ around 
$g_{m\phi^\dagger\phi} \simeq 10^{-7}$.
This is because for $g_{m\phi^\dagger\phi} \gtrsim 10^{-7}$, 
as is described in Section~\ref{Sec:Bol_sys}, the mediator 
is thermalized through the decay and inverse decay process, 
$m \leftrightarrow \phi \phi^\dagger$, 
and remains in chemical equilibrium with the SM sector until the 
final freeze-out of the  DM number density.
In this case, the DM relic density is insensitive to 
$g_{m\phi^\dagger\phi}$ and $g_{dd\phi^\dagger\phi}$, 
and is determined almost through the DM mass $m_d$, 
the DM-mediator mass ratio $r$ and 
the $d$-$m$ coupling $g_{ddmm}$.

\subsection{Temperature of $d$-$m$ system} 
\label{Sec:T}

\begin{figure}[h]
\begin{center}
\includegraphics[clip, width=87mm]{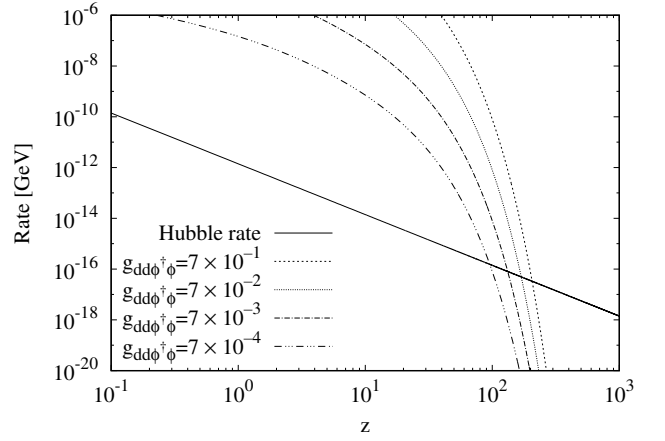}
\end{center}
\caption{Hubble rate $H$ and $d \phi \to d
\phi$ scattering rate $\langle \sigma v \rangle_{d
\phi \to d \phi} n_{\phi}$ for each $g_{dd\phi^\dagger\phi}$.
We take $m_{d} = 1\,\text{TeV}$.}
\label{Fig:KineticEqu}
\end{figure}

Throughout this work, we assume that the temperatures of 
$d$-$m$ system holds on the background temperature 
even after they decouple from the SM particles. 

Kinetic equilibrium of $d$ and the SM fields ensures that they 
evolve in a common background with a temperature, even if the 
chemical equilibrium of them is not achieved. 
Fig.~\ref{Fig:KineticEqu} shows the comparison of the Hubble 
expansion rate $H$ and the interaction rate of $d \phi 
\leftrightarrow d \phi$. 
We take $m_{d} = 1\,\text{TeV}$, $g_{ddmm} = 1$, and 
$g_{m\phi^\dagger\phi} = 1 \times 10^{-8}$. 
The interaction rate dominates $H$ in the region of 
$z \lesssim \mbox{a few} \times 10^{2}$.
For $z \gtrsim \mbox{a few} \times 10^{2}$, on the other hand,
the dark sector evolves in the temperature $T_{\rm dark}$
which may be different from the SM temperature $T_{\rm SM}$. 

In the example 1 shown in Sec.~\ref{Sec:evo}, 
$T_{\rm dark}=T_{\rm SM}$ does not hold after
the mediator life-time scale $z_{m\phi^\dagger\phi}$.
The two-step DM density decreasing profile (``terrace'' behavior)
is maintained even if we take into account the effects of
$T_{\rm dark} \ne T_{\rm SM}$, 
though we neglected the effects
in our numerical computations in this paper.
The issue will be discussed further in our future publication.

It is also possible to modify our toy model to keep $T_{\rm dark} = T_{\rm SM}$
for a longer period, assuming, e.g., the neutrino portal couplings for the 
dark sector instead of the Higgs portal coupling.

\section{A Realization in Pionic Dark Matter Scenario}
\label{sec:real-pion-dark}

As we have shown in the previous section, 
the $dd \leftrightarrow mm$ scattering amplitude needs to be 
strong enough to achieve the observed DM
relic density $\Omega_{\rm dm} h^2 = 0.1188 \pm 0.0010$
in our degenerated mediator setup.
If we assume the DM and mediator particles are elementary,
it is extremely difficult to obtain such a strong interaction
without conflicting with the Landau pole problem, however.
We here show both of
the degeneracy between the DM particle ($d$)
and the mediator particle ($m$), $m_{d} \simeq m_{m}$,
and the marginally strong interaction in $dd \leftrightarrow mm$
scattering can be accommodated in models of dark strong
dynamics having 
the dark pions~\cite{Bai:2010qg,Buckley:2012ky,Bhattacharya:2013kma}
as composite particles.
Note the dark pions exist ubiquitously in models of 
electroweak symmetry breaking, including
technicolor\cite{Ryttov:2008xe},
composite Higgs\cite{Frigerio:2012uc,Carmona:2015haa}
and also in classically scale invariant extentions of the 
SM\cite{Hur:2011sv,Ametani:2015jla,Hatanaka:2016rek,Holthausen:2013ota}.
Note also the 
SIMP mechanism~\cite{Hochberg:2014dra,Hochberg:2014kqa,Hochberg:2015vrg}
is embedded in the dark pion scenario, although the DM relic density decreasing
mechanism presented in this paper does not rely on it.

\subsection{Dark QCD and Dark Pions}
\label{sec:dark-qcd-dark}

We consider a model in which both DM particle $d$ and mediator $m$ are 
unified in a dark pion multiplet.
The dark pions are hypothetical pseudo Nambu-Goldstone bosons associated
with dynamical breaking of a newly introduced dark chiral
symmetry.
They often are the lightest BSM particle existing in models with
a dark strong dynamics (dark QCD),
and are regarded as the DM candidate particles.
In this section, we show the $d$-$m$ unification 
and the marginally strong $dd\leftrightarrow mm$ amplitude
can be achieved in a setup with the dark pions.

We introduce a new strong Yang-Mills gauge dynamics, termed ``dark QCD'' as, 
\begin{equation}
  {\cal L}_{\rm DQCD} = -\dfrac{1}{4g_{Ds}^2} G^a_{\mu\nu} G^{a\mu\nu}
                  +\bar{\psi} i\fsl{D} \psi,
\label{eq:darkqcd}
\end{equation}
in analog to the usual quantum chromodynamics (QCD).
Here the dark quark fermion field $\psi$ forms
a dark isospin doublet
\begin{equation}
  \psi_L = \left(
    \begin{array}{c}
      U_L \\
      D_L
    \end{array}
  \right),
  \qquad
  \psi_R = \left(
    \begin{array}{c}
      U_R \\
      D_R
    \end{array}
  \right),
\end{equation}
and belongs to the fundamental representation of the dark QCD gauge
group.
The fermion fields with left- and right-handed chiralities
are specified by using subscripts $L$ and $R$, respectively.
The Lagrangian Eq.(\ref{eq:darkqcd}) enjoys global 
$SU(2)_L\times SU(2)_R$ dark chiral symmetry,
\begin{eqnarray}
  & &
  \left(
  \begin{array}{c}
    U_L \\
    D_L
  \end{array}
  \right) \to
  \left(
  \begin{array}{c}
    U'_L \\
    D'_L
  \end{array}
  \right) = 
  \exp\left(i\sum_{a} \dfrac{\tau^a}{2} \theta_L^a\right)
  \left(
  \begin{array}{c}
    U_L \\
    D_L
  \end{array}
  \right),
\nonumber\\
& &
\\
  & &
  \left(
  \begin{array}{c}
    U_R \\
    D_R
  \end{array}
  \right) \to
  \left(
  \begin{array}{c}
    U'_R \\
    D'_R
  \end{array}
  \right) = 
  \exp\left(i\sum_{a} \dfrac{\tau^a}{2} \theta_R^a \right)
  \left(
  \begin{array}{c}
    U_R \\
    D_R
  \end{array}
  \right),
  \nonumber\\
  & &
\end{eqnarray}
with $\tau^a$ being the Pauli $SU(2)$ matrix.
In the Lagrangian Eq.(\ref{eq:darkqcd}), the dark gluon field $G_\mu^a$ couples
with the dark quark $\psi$ through the covariant derivative,
\begin{equation}
  D_\mu \psi = \partial_\mu \psi + i G_\mu^a T^a \psi,
\end{equation}
with $T^a$ being the fundamental representation matrix of the dark QCD gauge
symmetry.
The dark gluon field strength $G^a_{\mu\nu}$ is defined as usual
\begin{equation}
  G^a_{\mu\nu} T^a = \partial_\mu G^a_{\nu} T^a
                  -\partial_\nu G^a_{\mu} T^a
                  + i G^a_{\mu} G^b_{\nu} [T^a, T^b].
\end{equation}
Note here that both the dark fermion and the dark gluon are blind to 
the SM gauge group. These fields therefore contribute to the dark component
in the universe.

The negative beta function in the Yang-Mills theory 
renormalization group equations
makes the dark QCD gauge coupling strength
$g_{Ds}$ non-perturbatively strong and induces very strong attractive
force between $\psi$ and $\bar{\psi}$, which triggers a $\bar{\psi}\psi$
condensate
\begin{equation}
  \VEV{\bar{\psi}\psi} \ne 0  
\end{equation}
and dynamical breaking of the dark chiral symmetry,
\begin{equation}
  SU(2)_L \times SU(2)_R \to SU(2)_V,
\label{eq:symmetrybreaking}
\end{equation}
in a manner similar to the usual QCD\@.

It is now apparent how dark pions $\pi_D^a$ ($a=1,2,3$) appear
in this setup.  They are the Nambu-Goldstone bosons associated 
with the dynamical symmetry breaking Eq.(\ref{eq:symmetrybreaking}).
Due to the exact chiral symmetry, however, the dark pions
remain exactly massless in this model.
We therefore introduce explicit symmetry breaking terms,
\begin{equation}
  {\cal L}_{\rm explicit}
  = - \bar{\psi}_L m_\psi \left(
       1 - \dfrac{1}{\Lambda_s^2} \phi^\dagger \phi
         - \dfrac{1}{\Lambda_p^2} i\tau^3 \phi^\dagger \phi
      \right) \psi_R
      +{\rm h.c.},  
\label{eq:explicit}
\end{equation}
with $\phi$ denoting the $SU(2)_W$ doublet SM Higgs field.
The explicit breaking terms Eq.(\ref{eq:explicit}) make
the dark pions massive.
The dark pions interact with the SM Higgs field $\phi$
also through the explicit breaking terms Eq.(\ref{eq:explicit}).
Possible origin of these explicit symmetry violating terms 
Eq.(\ref{eq:explicit}) will be dealt in the next subsection
in a renormalizable field theory framework.
In this subsection, we concentrate on the impacts of these terms 
in the dark pion phenomenologies.

The dark pion low energy effective theory can be described by using the
chiral Lagrangian,
\begin{equation}
{\cal L}_{\chi} = \dfrac{f^2}{4} \tr \left[ 
    \partial_\mu U^\dagger \partial^\mu U
  \right] + 
  \dfrac{f^2}{4} \tr \left[
    \chi^\dagger U + U^\dagger \chi
  \right] ,
\label{eq:darkchirallag}
\end{equation}
with $f$ being the dark pion decay constant.
Here the non-linear chiral field $U$ is expressed using the 
dark pion field $\pi_D$,
\begin{equation}
  U = \exp\left(  \dfrac{i}{f} \sum_a \tau^a \pi_D^a
      \right).
\end{equation}
The effects of the explicit violation of 
the dark chiral symmetry, Eq.(\ref{eq:explicit}),
can be analyzed by using
\begin{equation}
  \chi = 2B m_\psi \left( 1 - \dfrac{1}{\Lambda_s^2} \phi^\dagger\phi
   +\dfrac{1}{\Lambda_p^2} i\tau^3 \phi^\dagger \phi
  \right),
\label{eq:explicitchiral}
\end{equation}
with $B$ being a low energy constant related with the 
dark quark pair condensate.

Expanding the chiral Lagrangian Eq.(\ref{eq:darkchirallag}) in
terms of the dark pion field $\pi_D$, we obtain
\begin{eqnarray}
  {\cal L}_\chi &=& 
  \frac{1}{2} \sum_a (\partial_\mu \pi_D^a)(\partial^\mu \pi_D^a)
 -\frac{1}{2} m_{\pi_D}^2 \sum_a \pi_D^a \pi_D^a
  \nonumber\\
  & & + g_{\pi_D \pi_D \phi^\dagger \phi} \sum_a \pi_D^a \pi_D^a \phi^\dagger \phi
  \nonumber\\
  & &
      + g_{\pi_D^3 \phi^\dagger \phi} m_{\pi_D} \pi_D^3 \phi^\dagger \phi
      + \cdots .
\end{eqnarray}
It should be noted here that all of the dark pions $\pi_D^1$, 
$\pi_D^2$, $\pi_D^3$ share the identical mass $m_{\pi_D}$, which
can be evaluated by using the low energy
constant $B$ and the dark quark mass $m_\psi$,
\begin{equation}
  m_{\pi_D}^2 = 2B m_\psi .
\end{equation}
We also find the dark pions couple with the SM Higgs boson through
Eq.(\ref{eq:explicit}) as,
\begin{equation}
  g_{\pi_D \pi_D \phi^\dagger \phi} = \dfrac{1}{2} \dfrac{m_{\pi_D}^2}{\Lambda_s^2},
\end{equation}
which plays an important role to thermalize the dark pions in the early
universe.

Note that the Lagrangians Eq.(\ref{eq:darkqcd}) and Eq.(\ref{eq:explicit})
are invariant under the fermion transformation,
\begin{equation}
  U_L \to -U_L, \qquad
  U_R \to -U_R,
\label{eq:symmetry}  
\end{equation}
which also survives as an exact symmetry even after the dynamical 
chiral symmetry breaking.
It is easy to see that 
$\pi_D^1$ and $\pi_D^2$ are odd under the transformation 
Eq.(\ref{eq:symmetry})
\begin{equation}
  \pi_D^1 \to -\pi_D^1, \qquad
  \pi_D^2 \to -\pi_D^2 .
\end{equation}
They are therefore stable and can be considered as the
DM candidate particles.
On the other hand, the third component of the dark pion $\pi_D^3$
is even under the symmetry Eq.(\ref{eq:symmetry}).
It then decays into the SM Higgs bosons through the coupling
\begin{equation}
  g_{\pi_D^3 \phi^\dagger \phi} = \dfrac{f m_{\pi_D}}{\Lambda_p^2} .
\end{equation}
We identify the third component dark pion $\pi_D^3$ 
as the mediator in the secluded DM scenario.
In this manner, the DM and the mediator particles are unified in the
same dark isospin multiplet.
The dark pion scattering amplitude can be evaluated 
by using the low energy theorem,
\begin{eqnarray}
  {\cal M}(\pi_D^1 \pi_D^1 \leftrightarrow \pi_D^3 \pi_D^3)
  &=& {\cal M}(\pi_D^2 \pi_D^2 \leftrightarrow \pi_D^3 \pi_D^3)
  \nonumber\\
  &=& \dfrac{s-m_{\pi_D}^2}{f^2} .
\label{eq:lowenergytheorem}
\end{eqnarray}
The amplitude Eq.(\ref{eq:lowenergytheorem}) is strong enough to
make the secluded DM scenario based on this setup.
If we deduce the $ddmm$ interaction coupling $g_{ddmm}$ of Eq.~(\ref{Eq:Lag_sce1})
from the $\pi_D^1\pi_D^1 \leftrightarrow \pi_D^3 \pi_D^3$ amplitude
at the threshold, we obtain
\begin{equation}
  4g_{ddmm} = \dfrac{3m_{\pi_D}^2}{f^2}.
\end{equation}
The marginally strong amplitude $g_{ddmm} \sim 1$ can thus be
easily achieved for the massive dark pions with $m_{\pi_D} \sim f$.
The other phenomenological couplings 
$g_{m\phi^\dagger\phi}$, 
$g_{dd\phi^\dagger\phi}$ and 
$g_{mm\phi^\dagger\phi}$ are
\begin{equation}
  g_{m\phi^\dagger\phi} = g_{\pi_D^3 \phi^\dagger\phi}, \qquad
  g_{dd\phi^\dagger\phi} = g_{mm\phi^\dagger\phi} = g_{\pi_D \pi_D \phi^\dagger \phi}.
\end{equation}

Note that the dark baryons also potentially contribute to 
the DM relic abundance in this model.  The dark baryon 
relic abundance turns out, however, to be negligibly small
in its minimal setup of the dark QCD at the TeV scale,
as demonstrated in the techni-baryon context in 
Ref.~\cite{Chivukula:1989qb}.
In models of asymmetric dark baryon DM, our mechanism
to decrease the dark pion number density can be applied 
to make the dark pions harmless in the 
cosmology\cite{Farina:2015uea,Freytsis:2016dgf}.

\subsection{A UV completion}

The purpose of this subsection is to give a possible renormalizable
UV completion behind the explicit breaking terms Eq.(\ref{eq:explicit}).
For such a purpose, we introduce real scalar fields $S$ and $P$, which 
interact with the dark fermions through
the Yukawa Lagrangian
\begin{equation}
  {\cal L}_{\rm Yukawa} = 
  - \bar{\psi}_L y (S+i\tau^3 P) \psi_R
  - \bar{\psi}_R y (S-i\tau^3 P) \psi_L,
\label{eq:yukawa}
\end{equation}
with $y$ being the Yukawa coupling strength.
Note that the Yukawa Lagrangian
violates explicitly the $SU(2)_L \times SU(2)_R$ symmetry down to 
$U(1)_L \times U(1)_R$.

We introduce kinetic and potential terms for these scalar field 
in a renormalizable manner,
\begin{equation}
  {\cal L}_{L\sigma M} = \dfrac{1}{2} (\partial_\mu S)^2
                      +\dfrac{1}{2} (\partial_\mu P)^2
                      -V(S, P),
\end{equation}
with
\begin{eqnarray}
  V &=& \dfrac{\lambda}{4} \left(
        S^2 + P^2 - F^2
      \right)^2 
    \nonumber\\
    & &
    + \dfrac{1}{4} \epsilon_\Delta F^2 (S^2-P^2)
    - \epsilon_S F^3 S  
    - \epsilon_P F^3 P ,
    \nonumber\\
    & &
\label{eq:pot}
\end{eqnarray}
with $F$ being a constant with a mass dimension.
We also introduce the interaction among $S$, $P$ and the
SM Higgs field $\phi$ as
\begin{equation}
  {\cal L}_{\phi^\dagger \phi} = -\dfrac{\lambda_M}{2} (S^2 + P^2) \left(
    \phi^\dagger\phi -\dfrac{v^2}{2}
  \right). 
\label{eq:SP-higgs}
\end{equation}
Note that the dimension 4 operators in the potential Eq.(\ref{eq:pot}) 
and the interaction with the SM Higgs Eq.(\ref{eq:SP-higgs})
respect the $U(1)$ symmetry among $S$ and $P$, while the operators
with lower dimensions violate the $U(1)$ symmetry.  
The dimension 2 operators respect the symmetries $S\to -S$, $P\to -P$, 
while the dimension 1 operators do not.
The potential is arranged to give non-vanishing vacuum expectation 
values (VEVs) for $S$ (and $P$) around the $F$ scale.
It is convenient if we parametrize VEVs as
\begin{equation}
  \VEV{S} = \bar{\varphi} \cos\bar{\theta}, \qquad
  \VEV{P} = \bar{\varphi} \sin\bar{\theta} .
\label{eq:vac}
\end{equation}
We also rewrite the explicit symmetry violating parameters
$\epsilon_S$ and $\epsilon_P$ using new parameters 
$|\epsilon|$ and $\theta_{\epsilon}$,
\begin{equation}
  \epsilon_S = |\epsilon| \cos\theta_{\epsilon}, \qquad
  \epsilon_P = |\epsilon| \sin\theta_{\epsilon} .
\end{equation}
Note here that, if we take $\theta_{\epsilon}=0$, the angle $\bar{\theta}$
is determined as
\begin{equation}
  \bar{\theta} = \dfrac{\pi}{2} n, 
\label{eq:vac-str}
\end{equation}
with $n$ taking an integer value.
Much involved vacuum structure than Eq.(\ref{eq:vac-str}) 
can be obtained if we take non-trivial value of $\theta_\epsilon$.
Especially, in this case, the mass eigenstates
arising from $S$ and $P$ fields do {\em not} necessary align to 
the VEV direction.
The misalignment between the VEVs and the particle mass eigenstates
causes interesting phenomenologies in this setup.

It is a bit tedious but straightforward to obtain the 
low energy effective theory by integrating out the $S$ and 
$P$ fields around the vacuum at the tree-level.
Due to the complex structure of the vacuum Eq.(\ref{eq:vac}),
the dark quarks acquire a complex mass in general, which can be rotated
away by performing an appropriate chiral rotation in the dark quark fields.

\begin{figure}
  \centering
  \includegraphics[width=87mm]{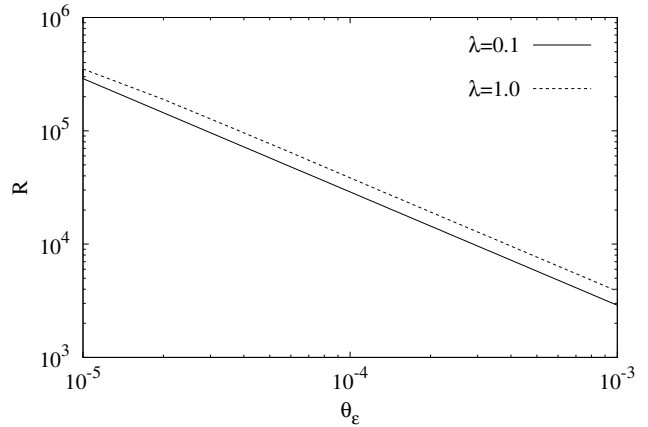}
%\\[10ex]
  \caption{The behaviour of Eq.(\ref{eq:fig7}).}
  \label{fig:fig7}
\end{figure}

Hereafter, we take
\begin{equation}
  \epsilon_\Delta = 0.01, \qquad |\epsilon| = 0.1, \qquad
  0 < \theta_\epsilon < \dfrac{\pi}{4}
\end{equation}
\begin{equation}
  \lambda \ge 0.1 . 
\end{equation}
Fig.~\ref{fig:fig7} shows the behaviour of 
\begin{equation}
  R=\dfrac{f}{m_{\pi_D}} \dfrac{g_{\pi_D\pi_D\phi^\dagger\phi}}{g_{\pi_D^3 \phi^\dagger \phi}} .
\label{eq:fig7}
\end{equation}
We see it is possible to take sufficiently large value of 
$g_{\pi\pi\phi^\dagger\phi}/g_{\pi_3 \phi^\dagger \phi}$, if we take the parameter
$\theta_\epsilon$ small enough.
Note that a $Z_2$ symmetry ($P\to -P$ symmetry) is restored
in the $\theta_\epsilon\to 0$ limit.
The smallness of $g_{m\phi^\dagger\phi}$ coupling 
($g_{\pi_D^3 \phi^\dagger\phi}$ coupling), as we assumed in 
our numerical demonstrations of Sec.~\ref{sec:numerical-results},
can thus be explained in a technically natural manner by the
smallness of the $\theta_\epsilon$ parameter.

\section{Summary and Outlook}
\label{sec:summary}

We have demonstrated in this article that the secluded DM scenario 
can successfully explain the observed relic DM density in the universe 
even in the case with the non-negligibly heavy mediator particle 
$m$ compared with the dark matter particle $d$ mass, $m_d \sim m_m$, if 
the mediator life-time is short enough and the $dd \leftrightarrow mm$ 
transition occurs rapidly enough.
The assumption $m_d \gg m_m$ imposed in the original secluded 
DM scenario\cite{Pospelov:2007mp,ArkaniHamed:2008qn} is therefore
not necessarily required. 
Allowing a heavy mediator having the mass $m_m$ nearly degenerate 
with the dark matter particle mass $m_d$, novel possibilities of
particle theory DM model buildings are now opened.
We gave a concrete renormalizable DM model, in which both the dark 
matter particle $d$
and the mediator particle $m$ are realized as the dark $SU(2)$ triplet
pseudo Nambu-Goldstone particles produced in the dark chiral symmetry 
breaking in the dark QCD.
The rapid transition $dd \leftrightarrow mm$ required in this
scenario is naturally achieved thanks to the compositeness of the
dark pions ($d$ and $m$) and the dark strong dynamics.

Although 
we concentrated on 
the computations
of the relic abundance in this manuscript, much work need to
be done in the DM phenomenologies in this scenario.
Due to the smallness of the DM coupling with the SM sector,
the direct detection of the secluded DM 
and the production of the DM particles in the high energy
collider experiments become rather challenging.
The large $dd\to mm$ amplitude and the subsequent decay
of the mediator particle $m$, on the other hand, induce
interesting signals in the indirect astrophysical DM detection 
experiments.
The dark pion model we proposed in this paper can easily
incorporate the DM particle number decreasing mechanism
through the $3\to 2$ scattering via the WZW interaction
(SIMP mechanism) in addition to the DM number decreasing
from the mediator decay.
These issues will be discussed in a separate publication.

\section*{Note added}
During the completion stage of this manuscript, two
papers\cite{Farina:2016llk,Dror:2016rxc}
which discuss massive mediator in the cannibal DM scenario
appeared in the arxiv.

\section*{Acknowledgements}
\label{sec:acknowledgements}

We thank Fumihiro Takayama, Paolo Gondolo and Shigeki Matsumoto
for useful discussions and valuable comments.
This work is supported in part by the JSPS Grant-in-Aid for 
Scientific Research 15K05047 (M.T.) and 16K05325 and 
16K17693 (M.Y.).

\appendix

\section{Terrace structure studied in a semi-analytic manner}
\label{Sec:ana_est}

In this appendix, we study the terrace structure we numerically
found in Sec.\ref{Sec:evo} more closely using an analytic method.
A semi-analytic formula to evaluate the DM relic abundance 
$Y_d^{\rm relic}$ is also given.
We assume the mediator $m$ and the DM $d$ degenerate in mass, 
$m_d=m_m$, as in Sec.\ref{Sec:evo}.
They also possess the same size couplings with the SM Higgs
field, $g_{mm\phi^\dagger\phi}=g_{dd\phi^\dagger\phi}$, thus
$\VEV{\sigma v}_{dd\leftrightarrow \phi^\dagger\phi}=
 \VEV{\sigma v}_{mm\leftrightarrow \phi^\dagger\phi}$.
The total number of relativistic degrees of freedom $g_*$ 
is taken to be $g_*=106.75$. 

We first consider the fake freeze-out, the decoupling of the
DM-mediator system from the SM particles.
Since the mediator decay process is not active at the
fake freeze-out time scale, we can neglect the 
$\VEV{\Gamma}_{m\leftrightarrow \phi^\dagger\phi}$ term in the Boltzmann equation
(\ref{Eq:Boltz_1_2}).
We also know $n_d\simeq n_m$, ($n_d^{\rm eq}=n_m^{\rm eq}$)
and the Boltzmann equation describing the fake freeze-out
behavior can be written as
\begin{equation}
  \dfrac{d n_{d+m}}{dt} + 3H n_{d+m}
  = -\frac{1}{2} \VEV{\sigma v}_{dd\leftrightarrow \phi^\dagger\phi}
     \left[
       (n_{d+m})^2 - (n_{d+m}^{\rm eq})^2
     \right],
\label{eq:Boltz_3}
\end{equation}
with
\begin{equation}
  n_{d+m} \equiv n_d + n_m, \qquad
  n_{d+m}^{\rm eq} = n_d^{\rm eq} + n_m^{\rm eq}.
\end{equation}
The Boltzmann equation (\ref{eq:Boltz_3}) can be converted to 
a form
\begin{equation}
  \dfrac{d}{dz} Y_{d+m} = - A_{\rm fake} z^{-n-2} \left[
    (Y_{d+m})^2 - (Y_{d+m}^{\rm eq})^2
  \right],
\label{eq:Boltz_3a}
\end{equation}
with 
\begin{math}
  z \equiv m_d/T.
\end{math}
$Y_{d+m}$ is defined as
\begin{equation}
   Y_{d+m} = Y_d + Y_m, \qquad
   Y_{d+m}^{\rm eq} = Y_{d}^{\rm eq} + Y_m^{\rm eq}.
\end{equation}
Here $Y_d$ and $Y_m$ are number densities normalized by entropy density $s$,
\begin{math}
  Y_d \equiv n_d/s
\end{math}
and
\begin{math}
  Y_m \equiv n_m/s.
\end{math}
The thermal equilibrium $Y_d$ and $Y_m$ are
\begin{equation}
  Y_d^{\rm eq} = Y_m^{\rm eq} = \dfrac{1}{2} a z^{3/2} e^{-z},
\end{equation}
with
\begin{equation}
  a \equiv \dfrac{45}{2\pi^4} \sqrt{\dfrac{\pi}{2}} \dfrac{1}{g_*}.
\end{equation}
We therefore obtain
\begin{equation}
  Y_{d+m}^{\rm eq} = az^{3/2} e^{-z}.
\end{equation}
The coefficient $A_{\rm fake}$ in Eq.(\ref{eq:Boltz_3a})
comes from the $dd\leftrightarrow \phi^\dagger\phi$ cross section,
\begin{eqnarray}
  A_{\rm fake} 
  &\equiv& \dfrac{1}{2} \left. \left[
      \dfrac{z \VEV{\sigma v}_{dd\leftrightarrow \phi^\dagger\phi} s}{H(T)}
  \right] \right|_{T=m_d}
  \nonumber\\
  &=& \sqrt{\dfrac{\pi}{45}{\displaystyle g_*}} m_d M_{\rm pl} 
      \sigma^{(0)}_{\rm fake}
\end{eqnarray}
Here $\sigma^{(0)}_{\rm fake}$ is defined through
\begin{equation}
  \dfrac{1}{2} \VEV{\sigma v}_{dd\leftrightarrow\phi^\dagger\phi}
  = \sigma_{\rm fake}^{(0)} \left(\dfrac{T}{m_d}\right)^n.
\end{equation}
Note that the $dd\leftrightarrow \phi^\dagger\phi$ process
($mm\leftrightarrow \phi^\dagger\phi$ process) occurs through the
$s$-wave.
We therefore use $n=0$ in our numerical estimates.

The freeze-out phenomenon in the type of Boltzmann equation 
(\ref{eq:Boltz_3a}) has been extensively studied in the 
textbook~\cite{Kolb:1990vq}.
We here only quote the results.
The time-scale at which $Y_{d+m}$ starts to exhibit the fake freeze-out
behavior ($z_{\rm fake}$)
can be defined by
\begin{equation}
  Y_{d+m}(z_{\rm fake}) - Y_{d+m}^{\rm eq}(z_{\rm fake})
  = c_{\rm fake} Y_{d+m}^{\rm eq}(z_{\rm fake}),
\end{equation}
with $c_{\rm fake}$ being an order $1$ constant.
It can be evaluated as
\begin{eqnarray}
  z_{\rm fake} &=& \ln \left[
                   (2+c_{\rm fake}) c_{\rm fake} A_{\rm fake} a 
                 \right]
             \nonumber\\
             & & - (n+1/2) \ln \left(
                    \ln \left[
                      (2+c_{\rm fake}) c_{\rm fake} A_{\rm fake} a 
                    \right]
                 \right).
    \nonumber\\
    & &
\end{eqnarray}
The textbook suggests $(2+c_{\rm fake}) c_{\rm fake} = n+1$ gives
the best fit.
Using this value of $c_{\rm fake}$, $n=0$ and the 
set of parameters in the evolution example 1, we obtain
\begin{equation}
  z_{\rm fake} \simeq 13.7,
\end{equation}
which agrees with the fake freeze-out time scale shown in 
Fig.~\ref{Fig:evolution_1}.

We next move to the final (true) freeze-out when the DM decouples
from the mediator.
Note that the mediator decay is already active at the age of the final
freeze-out.
The $\VEV{\Gamma}_{m\leftrightarrow \phi^\dagger\phi}$ term in Eq.(\ref{Eq:Boltz_1_2}) 
thus plays an important role.
On the other hand,
the $dd \leftrightarrow \phi^\dagger\phi$ and $mm\leftrightarrow \phi^\dagger\phi$
are negligibly small.
The Boltzmann equations can be approximated as
\begin{eqnarray}
  \dfrac{d}{dz} Y_d &=& -A_{ddmm} z^{-n'-2} \left[
                        (Y_d)^2 - (Y_m)^2
                        \right],
\label{eq:Boltz_4a}
  \\
  \dfrac{d}{dz} Y_m &=& -A_{ddmm} z^{-n'-2} \left[
                        (Y_m)^2 - (Y_d)^2
                        \right] - 4z B Y_m.
  \nonumber\\
  & &
\label{eq:Boltz_4b}
\end{eqnarray}
Here we used $\langle \Gamma \rangle_\text{ID} \ll \langle \Gamma \rangle_\text{D}$.
The coefficients $A_{ddmm}$ and $B$ are defined as
\begin{eqnarray}
  A_{ddmm} &=& \left. \left[
              \dfrac{z \VEV{\sigma v}_{dd\leftrightarrow mm} s}{H(T)}
            \right] \right|_{T=m_d}
  \nonumber\\
  &=& \sqrt{\dfrac{\pi}{45} {\displaystyle g_*}} m_d M_{\rm pl} 
      \sigma_{ddmm}^{(0)}.
  \\
  B&=& \dfrac{1}{8\pi} \sqrt{ \dfrac{45}{\pi g_*} } \Gamma_{m\to \phi^\dagger\phi} 
       \dfrac{M_{\rm pl}}{m_d^2}.
\end{eqnarray}
We define $\sigma^{(0)}_{ddmm}$ as
\begin{equation}
  \VEV{\sigma v}_{dd\leftrightarrow mm}
  = \sigma_{ddmm}^{(0)} \left(\dfrac{T}{m_d}\right)^{n'}.
\end{equation}
We should note here that $\sigma v$ in the $dd \leftrightarrow mm$ process 
depends on the velocity $v$ linearly in the case of
$d$-$m$ mass degeneracy.
The parameter $n'$ in the Boltzmann equations (\ref{eq:Boltz_4a})
and (\ref{eq:Boltz_4b}) should therefore be $n'=1/2$~\cite{Griest:1990kh}.

Summing up (\ref{eq:Boltz_4a}) and (\ref{eq:Boltz_4b}), we obtain
\begin{equation}
  \dfrac{d}{dz} Y_{d+m} = -4z B Y_m.
\label{eq:Boltz_4c}
\end{equation}
Note also that $Y_d$ tracks $Y_m$ very closely until the final
freeze-out $z_f$, and thus
\begin{equation}
  Y_d \simeq Y_m \simeq \dfrac{1}{2} Y_{d+m}.
\end{equation}
Eq.(\ref{eq:Boltz_4c}) can be solved as
\begin{equation}
  Y_d \simeq Y_m \simeq a' \exp\left( -Bz^2 \right),
\label{eq:Boltz_4d}
\end{equation}
for $z < z_f$.  
Here $a'$ denotes the integral constant.
We assume further that the behavior $Y_m \simeq a' \exp\left(-Bz^2\right)$
is valid even at $z\simeq z_f$
and solve the Boltzmann equation in the form of 
\begin{equation}
  \dfrac{d}{dz} Y_d = - A_{ddmm} z^{-n'-2} \left[
    (Y_d)^2 - (\tilde{Y}_d)^2 
  \right],
\label{eq:Boltz_5}
\end{equation}
with 
\begin{equation}
  \tilde{Y}_d \equiv a' \exp\left( - Bz^2 \right),
\end{equation}
instead of its original form (\ref{eq:Boltz_4a}).
The integral constant $a'$ is determined 
by fitting Eq.(\ref{eq:Boltz_4d}) with the numerical 
solution around $z \simeq z_f$.
It can also be determined roughly through matching with 
the $z_{\rm fake}$ epoch physics as we will show later.

The freeze-out phenomenon in Eq.(\ref{eq:Boltz_5}) can now be
analyzed in a manner similar to the textbook calculation of the
standard cold thermal relic abundance.
There are a couple of important differences in (\ref{eq:Boltz_5}), however:
the fractional $n'=1/2$ and the $\exp(-Bz^2)$ damping behavior of 
$\tilde{Y}_d$.
We see in below how these differences affect the freeze-out phenomenon 
in (\ref{eq:Boltz_5}).

We introduce 
\begin{equation}
  \Delta \equiv Y_d - \tilde{Y}_d,
\end{equation}
and define the freeze-out time scale $z_f$ by
\begin{equation}
  \Delta(z_f) = c_f \tilde{Y}_d(z_f),
\label{eq:def-freezeout}
\end{equation}
with $c_f$ being an order $1$ constant.
The Boltzmann equation (\ref{eq:Boltz_5}) can be expressed as
\begin{equation}
  \dfrac{d}{dz} \Delta
  = - \dfrac{d}{dz} \tilde{Y}_d
    - A_{ddmm} z^{-n'-2} \Delta (2\tilde{Y}_d+\Delta),
\label{eq:Boltz_5a}
\end{equation}
which can be solved approximately at $z=z_f$ as
\begin{eqnarray}
  \Delta(z_f) &\simeq& 
    - \dfrac{z_f^{n'+2}}{A_{ddmm}} 
      \left. 
        \dfrac{\dfrac{d}{dz} \tilde{Y}_d}{(2+c_f) \tilde{Y}_d}
      \right|_{z=z_f}
  \nonumber\\
  &=& \dfrac{2B}{A_{ddmm}} \dfrac{z_f^{n'+3}}{2+c_f}.
\label{eq:comp-freezeout}
\end{eqnarray}
Comparing Eq.(\ref{eq:comp-freezeout}) with 
Eq.(\ref{eq:def-freezeout}), we obtain
\begin{eqnarray}
  \dfrac{2B}{A_{ddmm}} \dfrac{z_f^{n'+3}}{2+c_f}
  = c_f a' \exp\left( -Bz_f^2 \right),
\label{eq:fo-scale0}
\end{eqnarray}
which leads to a formula to determine the freeze-out time
scale
\begin{eqnarray}
  z_f^2 &\simeq& 
    \dfrac{1}{B} \ln \left[
      \dfrac{(2+c_f)c_f}{2} \dfrac{A_{ddmm} a'}{B}
    \right]
  \nonumber\\
  & & -\dfrac{n'+3}{2} \dfrac{1}{B}
       \ln \left(
    \dfrac{1}{B} \ln \left[
      \dfrac{(2+c_f)c_f}{2} \dfrac{A_{ddmm} a'}{B}
    \right]
       \right)
\nonumber\\ 
& & + \cdots .
\label{eq:fo-scale}
\end{eqnarray}
Note $z_f^2$ is proportional to $\ln A_{ddmm}$ in Eq.(\ref{eq:fo-scale}).
This is in contrast to the usual cold thermal relic computation
in which $z_f$ is proportional to $\ln A$.
This property comes from the $\exp(-Bz^2)$ damping behavior
of $\tilde{Y}_d$ in this scenario.
We also note very slow convergence of the series expansion 
Eq.(\ref{eq:fo-scale}).
In our numerical anslysis, we therefore use Eq.(\ref{eq:fo-scale0}) 
directly, rather than Eq.(\ref{eq:fo-scale}).

Once we determine $z_f$, we can compute the relic abundance 
at $z\to \infty$ by
\begin{equation}
  \label{eq:2}
  Y_d^{\rm relic} = \lim_{z\to \infty} \Delta(z).
\end{equation}
For $z \gg z_f$, (\ref{eq:Boltz_5a}) can be approximated as
\begin{equation}
  \dfrac{d}{dz} \Delta 
  = - A_{ddmm} z^{-n'-2} \Delta^2.
\label{eq:Boltz_5b}
\end{equation}
Integrating (\ref{eq:Boltz_5b}) from $z_f$ to $\infty$, we obtain
\begin{equation}
  Y_d^{\rm relic} \simeq \dfrac{n'+1}{A_{ddmm}} z_f^{n'+1}.
\label{eq:Boltz_5c}
\end{equation}
Here the initial value uncertainty ($\Delta(z_f)$) is
absorbed in the uncertainty in $z_f$.
Note (\ref{eq:Boltz_5c}) is identical to the textbook formula
for the cold thermal relic abundance, except for the fractional
value of $n'=1/2$.
We obtain $a'\simeq 8.5\times 10^{-9}$ 
and $Y_d^{\rm relic}\simeq 4.2\times 10^{-13}$
in our numerical analysis presented in Sec.~\ref{Sec:evo}. 
Using these values, we see Eq.(\ref{eq:Boltz_5c}) as combined with
(\ref{eq:fo-scale}) gives the best fit
with
\begin{equation}
  c_f \simeq 1.1,
\end{equation}
which is perfectly consistent with the assumption we made on $c_f$:
it is an order $1$ constant.
The corresponding freeze-out $z_f$ is calculated as
\begin{equation}
  z_f \simeq 564.
\label{eq:zf_num}
\end{equation}
Again Eq.(\ref{eq:zf_num}) is 
consistent with Fig.~\ref{Fig:evolution_1}.

The final task we need to carry out is to make a relation
between the fake freeze-out $z_{\rm fake}$ and the
final freeze-out $z_f$.
This can be done by computing the coefficient $a'$ in (\ref{eq:Boltz_4d})
in terms of $z_{\rm fake}$.
Assuming the textbook formula
\begin{equation}
  Y_{d+m} = \dfrac{(n+1) z_{\rm fake}^{n+1}}{A_{\rm fake}}
\end{equation}
gives the abundance at $z=z_{\rm fake}$ in Eq.(\ref{eq:Boltz_4d}),
we see
\begin{equation}
  a' = \dfrac{(n+1) z_{\rm fake}^{n+1}}{2A_{\rm fake}} \exp\left(
        B z_{\rm fake}^2
      \right) .
\label{eq:aprime}
\end{equation}
Eq.(\ref{eq:aprime}) gives a result consistent with our numerical fit on $a'$ 
within 40\% uncertainty.

\end{document}